\documentclass[prd,12pt,reprint, twocolumn,noshowpacs,nofootinbib,amsmath,amssymb,superscriptaddress,preprintnumbers]{revtex4-1}

\usepackage{graphicx,color}
\usepackage{caption}
\usepackage{amssymb,amsmath,mathrsfs}
\usepackage{cancel}
\usepackage{subfig}
\usepackage{hyperref}
\usepackage{verbatim,slashed}

\newcommand{\GeV}{\mathrm{GeV}}

\newcommand{\beq}{\begin{equation}}
\newcommand{\eeq}{\end{equation}}
\newcommand{\be}{\begin{eqnarray}}
\newcommand{\ee}{\end{eqnarray}}

\newcommand{\mm}{\mathrm{mm}}

\newcommand{\Ap}{A^\prime}
\newcommand{\hd}{h_{\rm D}}
\newcommand{\Hd}{H_{\rm D}}
\newcommand{\ed}{e_{\rm D}}
\newcommand{\ad}{\alpha_{\rm D}}
\newcommand{\vd}{v_{\rm D}}
\newcommand{\lamd}{\lambda_{\rm D}}
\newcommand{\mZ}{m_Z}
\newcommand{\mHd}{m_{h_{\rm D}}}
\newcommand{\mAp}{m_{A^\prime}}
\newcommand{\fb}{\mathrm{fb}}
\newcommand{\pT}{p_{\mathrm{T}}}
\newcommand{\UD}{U(1)_{\rm D}}
\newcommand{\thh}{\theta_h}

\def\lsim{\mathrel{\rlap{\lower4pt\hbox{\hskip1pt$\sim$}}
    \raise1pt\hbox{$<$}}}
\def\gsim{\mathrel{\rlap{\lower4pt\hbox{\hskip1pt$\sim$}}
    \raise1pt\hbox{$>$}}}

\begin{document}
\title{Rare $Z$ Boson Decays to a Hidden Sector}
\author{Nikita Blinov}
\affiliation{SLAC National Accelerator Laboratory, 2575 Sand Hill Rd., Menlo Park, CA 94025, USA}
\author{Eder Izaguirre}
\affiliation{Brookhaven National Laboratory, Upton, NY 11973, USA}
\author{Brian Shuve}
\affiliation{Harvey Mudd College, 301 Platt Blvd., Claremont, CA 91711, USA}
\affiliation{SLAC National Accelerator Laboratory, 2575 Sand Hill Rd., Menlo Park, CA 94025, USA}
\date{\today}

\preprint{SLAC-PUB-17159}

\begin{abstract}
We demonstrate that rare decays of the Standard Model $Z$ boson can be used to discover and characterize the nature of new hidden-sector particles. We propose new searches for these particles in soft, high-multiplicity leptonic final states at the Large Hadron Collider. The proposed searches are sensitive to low-mass particles produced in $Z$ decays, and we argue that these striking signatures can shed light on the hidden-sector couplings and  mechanism for mass generation.
\end{abstract}
\maketitle
  \section{Introduction}
  \label{sec:introduction}

  The Large Hadron Collider (LHC) is now probing the Standard Model
  (SM) and its possible extensions at energies exceeding the electroweak scale.
  The lack of any definitive discoveries beyond the Higgs boson, however, suggests that new
  physics within the reach of the LHC may appear in unexpected places. It is now
  widely appreciated that some classes of signatures are challenging
  to probe at a hadron collider. These include new electroweak multiplets with masses near the 
  electroweak scale~\cite{Chen:1995yu,Thomas:1998wy,Han:2013usa,Low:2014cba}; particles produced via the strong interactions that decay to approximately degenerate states \cite{Alwall:2008va}; ``stealth'' spectra~\cite{Fan:2011yu};
  or all-hadronic final states~\cite{Evans:2012bf}.

  In this paper, we focus on searches for low-mass particles, which can likewise be difficult to find  at the LHC. The signatures of new light particles are manifold and  are sensitive to both the manner of production (\emph{i.e.}, whether the low-mass states are produced directly or in the decays of heavier particles), as well as whether there exists a single new
  particle~\cite{Boehm:2003hm,Pospelov:2008zw} or an entire hidden sector or ``hidden
  valley''~\cite{Strassler:2006im,Strassler:2006ri,Han:2007ae}. When the new low-mass particles decay back into
   SM final states, the result is typically a high-multiplicity final
  state~\cite{Strassler:2006im,Han:2007ae,Batell:2009yf,Batell:2009jf}. Since the energy in the event is divided among
  many particles, the final-state particles are typically soft and could
  be lost in the enormous multijet backgrounds or contaminated by
  pile-up. However, hidden sectors can also give rise to spectacular signatures
  such as high multiplicities of leptons, displaced vertices, and other
  non-standard phenomena~\cite{Strassler:2006im,Strassler:2006ri,Han:2007ae,ArkaniHamed:2008qp, Cheung:2009su, Baumgart:2009tn}. These can be used to suppress the sizable SM backgrounds, provided the signals are sufficiently energetic to be efficiently reconstructed.

Due to its high luminosity, the LHC is an exquisite tool for
  studying hidden-sector particles that are too heavy to be produced in other intensity-frontier
  experiments. Through Phase II, the LHC  will produce
  unprecedented numbers of electroweak and Higgs bosons, exceeding the statistics
  from LEP by many orders of magnitude. By looking for new states produced in the rare decays of
  SM particles,  we may probe ever-smaller
  couplings between SM particles and hidden sectors. The production of hidden-sector
  particles in the decays of the SM Higgs boson has been well studied (see, for example, Ref.~\cite{Curtin:2013fra}). In 
  this study, we focus on rare $Z$ boson decays, which are most important in vector-portal models of hidden sectors.

  To achieve optimal sensitivity to hidden sectors at the LHC,
it is essential that the experiments are able to efficiently trigger on and
  reconstruct new low-mass objects. Dedicated search and
  reconstruction strategies are needed to uncover evidence for low-mass
  particles, especially if the particles are boosted and fail conventional isolation criteria. 
  Such strategies have been successfully implemented in several
  contexts, such as for lepton
  jets~\cite{Aad:2014yea,Aad:2015sms,Khachatryan:2015wka}; compressed
  stops~\cite{ATLAS:2017dnw,CMS:2017odo};  long-lived, low-mass particles decaying
  in the ATLAS hadronic calorimeter and muon spectrometer~\cite{Aad:2015asa}; and
  searches by the LHCb collaboration for long-lived particles  in the
  forward direction~\cite{Aaij:2016xmb}. However, comprehensive studies are
  needed to ensure that signals are not lost at trigger or event reconstruction
  level, and we undertake such a study with a focus on high-multiplicity (6+ particle) decays of 
the $Z$ boson into hidden sector particles.

\begin{figure}
    \centering
    \includegraphics[width=0.35\textwidth]{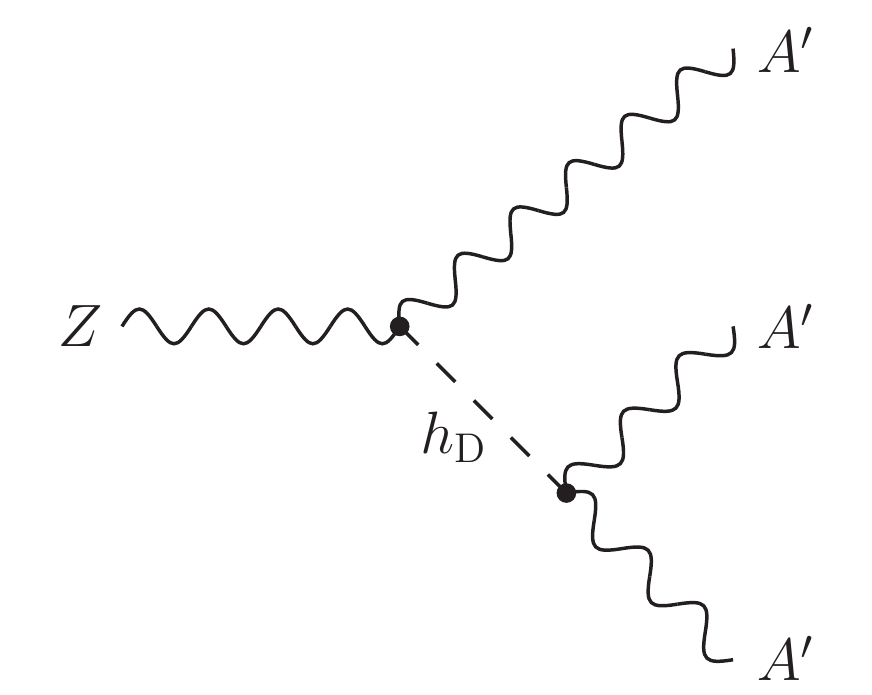}
    \caption{Feynman diagram illustrating dark Higgs ($\hd$) and dark photon ($A'$) production in rare $Z$ boson decays. 
      The dark photons in $\hd$ decay can be on- or off-shell.
    \label{fig:feynman_Zdecay}}
  \end{figure}

  We take as a concrete example a minimal hidden sector with a spontaneously broken Abelian gauge interaction 
\cite{Holdom:1985ag,Galison:1983pa}.
  The new gauge boson (the dark photon, $A'$), couples to the SM via kinetic mixing with the
  hypercharge boson, resulting in $Z$ decays into the hidden sector. We show that $Z$
  decays into $A'$ in association with the dark Higgs boson, $\hd$, lead to
  striking signatures such as multiple resonances, high multiplicities of soft
  leptons, and long-lived particles. The decay $Z\rightarrow A'\hd$ is a dark
  version of the Higgs-strahlung process and additionally allows for a direct test of the
  mass generation mechanism in the hidden sector. If the dark Higgs decays via
  $\hd\rightarrow A'A'$ as illustrated in Fig.~\ref{fig:feynman_Zdecay}, then
  $A'$ decays into SM particles can yield a final state with as many as six
  leptons. 

  We show that searches for dark Higgs-strahlung in rare $Z$ decays can
  be competitive with and exceed the sensitivity of existing and proposed search
  strategies for dark gauge and Higgs bosons, while also yielding new information.  Because the decay $Z\rightarrow A' \hd$ 
  is sensitive to the hidden-sector gauge coupling, $\ad$, a discovery of a signal in
  this process provides complementary information to direct searches for $A'$. Previous proposals entailed looking for $A'$ at the LHC via its contribution to Drell-Yan
  production above 10 GeV \cite{Hoenig:2014dsa,Ilten:2016tkc}, in rare 3-body $Z$ decays into leptons and a dark gauge boson \cite{Elahi:2015vzh}, as well as in decays of the SM Higgs boson \cite{Curtin:2013fra, Davoudiasl:2013aya, Curtin:2014cca, Chou:2016lxi}. Finally, Ref.~\cite{Liu:2017lpo} evaluated the prospects for discovering $A'$ and $\hd$ via Higgs-strahlung at future lepton colliders, where they focused on decays to invisible hidden-sector states.

  While our study is phenomenologically driven, we demonstrate sensitivity to
  models of hidden sectors that are well motivated by various shortcomings
  of the SM, especially the need to account for dark matter but also potentially outstanding
  problems in neutrino physics and other areas. Indeed, low-mass dark matter
  scenarios are typically only viable with additional low-mass mediators in the
  hidden sector~\cite{Lee:1977ua,Boehm:2003hm,Pospelov:2007mp}. Hidden sectors can also generically arise in ultraviolet
  completions of the SM such as string theory~\cite{Cvetic:1996mf,Abel:2008ai,Goodsell:2009xc}.

  We outline our benchmark hidden-sector model in Section~\ref{sec:model}. 
  We then enumerate the signatures of rare $Z$ decay into the
  hidden sector, focusing on signals with high lepton multiplicities and
  hidden-sector resonances. We give projected LHC sensitivities to prompt
  hidden-sector signals in Section \ref{sec:prompt} and  we discuss
  displaced signals in Section \ref{sec:displaced}. Our outlook is given in
  Sec.~\ref{sec:discussion}.

  \section{An Abelian Hidden Sector}\label{sec:model}

The benchmark model we consider is one of the simplest examples of a hidden sector:~a minimal $\UD$ gauge interaction spontaneously broken by a non-decoupled Higgs field. The model is specified by the following Lagrangian:
  \be
  \mathscr{L} &\supset& -\frac{1}{4}F^\prime_{\mu\nu}F^{\prime \mu\nu}+ \frac{\varepsilon_Y}{2} F'_{\mu\nu} B^{\mu\nu}\nonumber\\
  &&{} 
  + |\partial_\mu \Hd-i\ed A'_\mu \Hd|^2 - V(H,\Hd),
  \label{eq:lagrangian}
  \ee
  where $F^{\prime \mu\nu}$ ($B^{\mu\nu}$) is the $\UD$ (hypercharge) field
  strength, $\ed = \sqrt{4\pi \ad}$ is the $\UD$ gauge coupling, and $V$ is the scalar potential
  \be
  V(H,\Hd) &=& -\mu_H^2|H|^2  -\mu_{\Hd}^2 |\Hd|^2+ \kappa |H|^2|\Hd|^2\nonumber\\
  &&{} + \lambda |H|^4 + \lamd |\Hd|^4 .
  \label{eq:potential}
  \ee
  The SM and dark Higgs acquire vacuum expectation values (VEVs) $\langle H \rangle = (0,v/\sqrt{2})$ and $\langle \Hd\rangle = \vd/\sqrt{2}$, 
  breaking the electroweak and $\UD$ gauge symmetries, respectively. As a result, gauge eigenstates undergo mixing \cite{Holdom:1985ag}. For a recent review of the model, see Ref.~\cite{Curtin:2014cca}.

  The kinetic and mass terms for the gauge bosons can be simultaneously 
  diagonalized using the transformation (to leading order in $\varepsilon_Y$ and $\theta_Z$)
  \begin{eqnarray}
     Z_\mu& \rightarrow& Z_\mu - ( \theta_Z + \varepsilon \tan\theta_{\rm W})A'_\mu,\\
    A'_\mu &\rightarrow & A'_\mu + { \theta_Z} Z_\mu ,\\
    A_\mu &\rightarrow& A_\mu + \varepsilon \,A'_\mu,
  \end{eqnarray}
  where $\varepsilon \equiv \varepsilon_Y\cos\theta_{\rm W}$,  $\theta_{\rm W}$ is the weak mixing angle, and 
  \beq
  \theta_Z = { -} \frac{\varepsilon \tan\theta_{\rm W}\,{m_Z^2}}{\mZ^2-\mAp^2} + \mathcal{O}(\varepsilon^3)
  \eeq
  is the gauge-boson mixing angle. The $Z$ and $A'$ bosons have masses equal to the unshifted values at $\mathcal{O}(\varepsilon)$,  while the photon $A$ remains exactly massless.

  The scalar states can also undergo mixing after gauge-symmetry breaking. The mass eigenstates are
  \beq
  \begin{pmatrix}
    h \\
    h_{\rm D}
  \end{pmatrix} = 
  \begin{pmatrix}
    \cos\thh &  -\sin\thh \\  
    \sin \thh & \cos\thh
  \end{pmatrix}
  \begin{pmatrix}
    h^{(0)} \\
    h_D^{(0)}
  \end{pmatrix},
  \eeq
  where $h^{(0)}, \hd^{(0)}$ are the  $CP$-even gauge eigenstate components of
  $H$ and $\Hd$, respectively. In the limit of small mixed quartic coupling
  $\kappa$, the $h-\hd$ mixing angle is given by 
  \beq
  \sin \thh \approx \frac{\kappa}{2}\frac{v v_{\rm D}}{\lambda_{\rm D} v_{\rm D}^2 - \lambda v^2}.
  \eeq
  Except where otherwise noted, we assume  that the dominant
  hidden-sector portal to the SM is via the coupling between gauge bosons, $\varepsilon$.

  Because the $\UD$ gauge symmetry is spontaneously broken, the masses of
  hidden-sector particles are related to the symmetry-breaking parameter. In
  particular, the dark Higgs VEV gives a mass to the dark gauge boson, leading to
  a dark Higgs-strahlung  $\hd-A'-A'$ vertex by analogy with the symmetry
  breaking pattern in the SM. After the vector and scalar eigenstates mix,
  we obtain the following mass-basis Lagrangian: 
  \beq
  \mathscr{L} \supset  g_{\hd \Ap Z}\, \hd \Ap_\mu Z^\mu + g_{\hd \Ap \Ap}\, \hd \Ap_\mu {\Ap}^\mu + g_{\Ap \bar f f}\,\Ap_\mu\, \bar f\gamma^\mu f,
  \label{eq:interaction_lagrangian}
  \eeq
  along with additional terms that are not relevant for the phenomenology we study. The approximate couplings may be expressed simply in the $\mAp\ll \mZ$ limit,
  \begin{align}
    g_{h_{\rm D} \Ap Z} & ={2 {\ed} \theta_Z\cos \thh  \mAp}, \\
  g_{h_{\rm D} \Ap \Ap} & = \ed^2 v_{\rm D}, \\
  g_{\Ap \bar f f} & = \varepsilon e Q_f\label{eq:ap_current}. 
  \end{align}
  The first term in Eq.~\eqref{eq:interaction_lagrangian} gives rise to  $Z$ boson decay into $\hd$ and $\Ap$ with rate
  \be
  \Gamma_{Z\rightarrow \Ap \hd} = 
  \frac{2\ad { \,\theta_Z^2 \,\cos^2 \thh} \,\mAp^2}{3 \mZ}
  \left[1+\,\,\,\,\,\,\,\,\,\,\,\,\,\,\,\,\,\,\,\,\,\,\,\,\,\,\,\,\,\right.\nonumber\\
  \left.\frac{(\mZ^2+\mAp^2-\mHd^2)^2}{8\mZ^2 \mAp^2}\right]
  \beta\left(\frac{\mAp}{\mZ},\frac{\mHd}{\mZ}\right),\label{eq:z_branching}
  \ee
  where $\beta(x,y) = [(1-(x-y)^2)(1-(x+y)^2)]^{1/2}$. This rate vanishes as
  $m_{A'}\rightarrow0$, but becomes appreciable as $m_{A'}$ increases above
  $\sim10$ GeV, which is precisely where constraints  from low-energy colliders
   become ineffective.

  The decays of the $A'$ and the $\hd$ depend on the spectrum of the hidden
  sector. When $A'$ and $\hd$ are the lightest states,  they decay back into
  SM particles. The $A'$ decays dominantly into electrically charged SM fermion
  pairs via the coupling in Eq.~\eqref{eq:ap_current}. When kinematically
  allowed, the dark Higgs can decay into one or two on-shell $A'$ via
  $\hd\rightarrow A'{A'}^{(*)}$, and the $A'$ in turn decays into SM fermions.
  Finally, if $\mHd<m_{A'}$, then the $\hd$ can decay via two off-shell $A'$,
  although radiative corrections typically induce a comparable decay via scalar
  mixing.

  Having defined the model, we can consider new search strategies for discovering
  the hidden sector through the process, $Z\rightarrow
  A'\hd$. The specific signatures depend  on the mass hierarchy of the $A'$ and $\hd$. For
  $m_{A'}<\mHd$, the decay $\hd\rightarrow A'{A'}^{(*)}$ is kinematically allowed
  and occurs relatively rapidly, giving rise to prompt signals. We propose
  searches for such prompt signatures in Section~\ref{sec:prompt}. If instead
  $m_{A'}>\mHd$, then the $\hd$ lifetime is typically long; we explore this
  scenario in Section~\ref{sec:displaced}. Before we study the $Z\rightarrow \Ap \hd$ signal,
   however, we first  summarize existing constraints on the model.

  \subsection{Existing Constraints}
We focus on masses $m_{A'},\,\mHd\gtrsim1$ GeV which are most
  relevant for the LHC.\\

  {\noindent \bf Low-Energy $e^+e^-$ Colliders}:~The clean environment and high
  luminosities of low-energy $e^+e^-$ colliders allow for powerful searches for
  $A'$ and $\hd$ for masses below 10
  GeV. The kinetic-mixing coupling of $A'$ to fermions is studied in the
  radiative return reaction $e^+e^-\rightarrow A'\gamma$, $A'\rightarrow
  \ell^+\ell^-$ for electron and muon final states. In particular, the analysis
  from Ref.~\cite{Lees:2014xha} puts mass-dependent upper bounds on $\varepsilon$ in
  the range $3\times10^{-4}-10^{-3}$ for masses $m_{A'}\lesssim10$ GeV. BaBar and
  Belle also constrain hidden sectors through searches for the dark
  Higgs-strahlung process, $e^+e^-\rightarrow A' \hd$
  \cite{Lees:2012ra,TheBelle:2015mwa}. These searches constrain the combination
  of couplings $\ad \varepsilon^2\lesssim10^{-9}$, with the precise limits
  depending on $m_{A'}$ and $\mHd$. Belle II is expected to
  significantly extend this sensitivity to hidden sectors below 10 GeV, although
  in all cases $B$-factories lose sensitivity for $m_{A'}>10 $
  GeV.\\

  {\noindent \bf Electroweak Precision Observables (EWPO)}: Precision tests of
  electroweak symmetry breaking set stringent bounds for $\mAp < m_Z$.
  Bounds from EWPO
  arise predominantly from the mass mixing of $A'$ with $Z$, although
  other EWPO observables get modified by virtue of the fact that electrically
  charged SM particles acquire an effective millicharge under the $A'$ (see for
  instance, Ref.~\cite{Hook:2010tw} and a newer analysis including prospects for
  future improvements in EWPO at high-energy lepton colliders by
  Ref.~\cite{Curtin:2014cca}).\\
    
  {\noindent \bf  LHC Constraints}: The $A'$ mediates new contributions
  to Drell-Yan production of leptons at the LHC. A reinterpretation of a CMS
  study of the Drell-Yan process at $\sqrt{s} = 7$ TeV \cite{Chatrchyan:2013tia}
  sets constraints on the $A'$ mass via a dilepton resonance search
  \cite{Hoenig:2014dsa} (see Ref.~\cite{Hoenig:2014dsa} and an updated analysis
  by Ref.~\cite{Curtin:2014cca} on future prospects for Drell-Yan sensitivity to
  $A'$). A recent LHCb search for inclusive Drell-Yan production of dark photons 
  has set powerful constraints in the $m_{A'}\sim10-70$ GeV mass range 
  with 1.5 $\fb^{-1}$ of data at $\sqrt{s}=13$ TeV~\cite{Aaij:2017rft}.
   In the case where $\hd$ has an appreciable mixing with $h_{\rm
  SM}$, there are also constraints on the hidden sector via $h_{\rm
  SM}\rightarrow A'A'$ \cite{Curtin:2014cca,Aad:2015sva}; in the limit of small
  scalar mixing that is our focus, the latter searches do not apply any meaningful constraint.

\section{Prompt Multilepton Signals from Rare $Z$ Decays}\label{sec:prompt}

We have argued that the dark sector benchmark model in
Eq.~\eqref{eq:lagrangian} gives rise to striking signatures in the decay of the
SM $Z$ boson via $Z \rightarrow A'\hd$. Studying this process would allow for
not only the discovery of $A'$ and $\hd$, but can also provide information
about the hidden sector couplings and mass generation mechanisms. We now
examine in detail the phenomenology of this rare $Z$ decay.\\

\noindent {\bf Fully On-Shell Decays:}~If $\mHd > 2 \mAp$, then the following decay chain occurs:
\beq
p p \rightarrow Z \rightarrow A' \hd \rightarrow A' A' A'.
\eeq
The $A'$ can decay to various SM final particles. In principle, this gives
final-state signatures with three $A'$ resonances, as well as the intermediate
$\hd$ resonance. In the case of fully leptonic decays, this is an extremely
clean and spectacular signature. In practice, other $Z$ decays may give better
model sensitivity because of the small probability that
each $\Ap$ decays leptonically, as well the efficiency
penalties associated with reconstructing all six soft leptons. Combinatorics 
 may also present a challenge to reconstructing the signal. However,
even the more inclusive case with $Z\rightarrow \hd A'$, $\hd \rightarrow
4\ell$ (with the other $A'$ decaying to arbitrary final states) features
 two leptonically-decaying $A'$ resonances that together reconstruct the $\hd$.
   We therefore
investigate signatures with at least four leptons from the $\hd$ decay.

Due to the limited available phase space in six-body $Z$ decays, it is beneficial to use
multilepton triggers with low $\pT$ thresholds. The current triggers that best fit these
requirements are the three-lepton triggers.  For ATLAS, these
are~\cite{ATL-DAQ-PUB-2017-001}
\begin{itemize}
  \item three loose $e$'s: $\pT \geq 15,\; 8,\; 8$ GeV at L1 (17, 10, 10 at HLT),
  \item three $\mu$'s: $\pT > 6~\GeV$ ($3\times6$ at HLT).
\end{itemize}
For CMS,
we examine the thresholds from a multilepton analysis such as $ZZ\rightarrow4\ell$~\cite{CMS:2014xja}, which used 
\begin{itemize}
  \item three $e$'s: $\pT \geq 15,\; 8,\; 5$ GeV.
\end{itemize}
While this analysis was conducted at $\sqrt{s}=8$ TeV, the CMS dilepton trigger
thresholds did not increase appreciably between 8 TeV and 13 TeV
\cite{Khachatryan:2014qwa,Sirunyan:2017lae}, and we use this as a  proxy
for what can currently be expected in terms of trigger thresholds. We note that
trilepton thresholds may go up in the high-luminosity phase of the LHC;
however, in this case 4-lepton triggers could alternatively be used to keep $\pT$
thresholds low.\\

\noindent {\bf Partially off-shell decays:}~When $\mAp<\mHd<2\mAp$, the decay
$\hd\rightarrow A'A'$ is kinematically forbidden, but the semi-off-shell decay
$\hd\rightarrow A'{A'}^*\rightarrow A'\bar{f}f$ (where $f$ is a SM fermion) can occur. While $\Gamma_{\hd}$
is suppressed by a factor of $\varepsilon^2$ for semi-off-shell decays compared
to the case of fully on-shell decays, we find that the decay still occurs
promptly for the parameters of relevance to the LHC. Thus, the reaction proceeds as
follows:
\beq
p p \rightarrow Z \rightarrow A' \hd \rightarrow A' A' \bar{f}f.
\eeq
The principal difference between the semi-off-shell and fully on-shell decays 
 is that  one of the pairs of SM particles no longer
reconstructs a resonance. As we shall soon see, however, the backgrounds
are sufficiently low that we do not necessarily 
need to impose a $\mAp$ resonance reconstruction requirement on the
leptons within $\hd\rightarrow4\ell$ decays, and so the fully and partially on-shell $\hd$
decays can be studied simultaneously. Depending on how actual experimental conditions
compare to Monte Carlo simulations, these two cases may need to be studied separately. \\

\noindent {\bf Simulations:}~In performing our analysis, we generate parton-level events at leading order
with \texttt{MadGraph5\_aMC@NLO}~\cite{Alwall:2014hca}. To capture the effects
of initial-state radiation on signal acceptance, we generate
events with up to one extra parton and match them to the showered events using the
shower-$k_{\rm T}$ scheme~\cite{Alwall:2008qv}. We use the parton shower program
\texttt{Pythia 8.2}~\cite{Sjostrand:2006za,Sjostrand:2007gs,Sjostrand:2014zea}.\\

\noindent {\bf Reinterpretation of Existing Searches:}~Existing searches by
ATLAS and CMS are already sensitive to signals with high lepton multiplicities.
For example, CMS has a search for new electroweak supersymmetric (SUSY)  particles that decay to
three or more leptons \cite{Sirunyan:2017lae}. We re-interpret the results of
the low-$\slashed{E}_{\rm T}$, 4-lepton signal regions in terms of our signal
model. In order to derive the constraints from the CMS search, we must apply
lepton identification efficiencies, which are somewhat small for leptons with
low $p_{\rm T}$. Because Ref.~\cite{Sirunyan:2017lae} only provides the
low-$p_{\rm T}$ lepton tagging efficiencies for the most pessimistic working
point, we must use the pessimistic values and obtain a conservative result. The
true signal efficiency is almost certainly better than what we find, because
Ref.~\cite{Sirunyan:2017lae} states that a looser set of lepton identification
criteria are used for searches with four leptons, but does not specifically
state what these efficiencies are.

We find that the Signal Region (SR) H of Ref.~\cite{Sirunyan:2017lae}, which
requires 4 leptons and fewer than two opposite-sign, same-flavor (OSSF) lepton
pairs, is most sensitive to the hidden sector topology we study. In this case,
the $Z\rightarrow A'\hd\rightarrow 6\ell$  decay can pass the CMS signal
selections if two of the leptons are lost and only one OSSF pair is
reconstructed.\footnote{There is also a signal region with two OSSF pairs (SR G)~\cite{Sirunyan:2017lae}, 
which features a larger signal acceptance but
significantly  larger backgrounds (and systematic uncertainties). As a result,
the  signal sensitivity is better for SR H.} Using the $\mathrm{CL}_{\rm s}$
method \cite{Read:2002hq}, we estimate a constraint on the dark photon kinetic
mixing, $\varepsilon$, at the 95\% confidence level (c.l.).

The limits from the CMS multilepton search are better than the constraints from
EWPO, but somewhat worse than the recent bounds from LHCb \cite{Aaij:2017rft},
and so we do not show them explicitly. However, as argued above our
re-interpretation is almost certainly more conservative than the actual
analysis, and the existing CMS search may in fact place limits on the dark
Abelian Higgs model that are competitive with LHCb. Crucially, the CMS search
is not optimized for the hidden-sector signal, and so the fact that it is
already somewhat competitive with existing dedicated searches for $\Ap$ affirms
the important role of rare $Z$ decays in probing hidden sectors.\\

\noindent {\bf Proposal for New Multilepton Search:}~Our signal features
distinctive kinematics and multiple resonances in $\hd\rightarrow
A'A'\rightarrow4\ell$ decays. These features  can be exploited to significantly
reduce both the background and the uncertainty on its estimate compared to the
CMS SUSY signal regions while maximizing signal efficiency.

 We define a signal  region for the decay mode $Z\rightarrow A'\hd$, $\hd\rightarrow A'A'\rightarrow 4\ell$, while remaining agnostic about the third $A'$ decay mode. The selections for our signal region are:
\begin{itemize}
  \item \emph{Trigger selection:}~Require four muons with $\pT > 7\;\GeV$ or any four leptons with $\pT >15,\; 8,\; 7,\; 5$ GeV. All leptons must be in the central part of the detector ($|\eta|<2.5$).
\item \emph{Isolation:}~Require that each lepton be isolated from hadronic activity or photons. We do this by finding the scalar sum $p_{\rm T}$ of all hadrons or photons within $\Delta R = 0.3$ of the lepton and requiring that the sum be \emph{either} less than 5 GeV \emph{or} less than 20\% of the lepton $p_{\rm T}$.
\item \emph{$A'$ Signal Selection:}~Require at least 2 OSSF lepton pairs (each pair forms an $A'$ candidate).
  \item \emph{Suppression of $Z$ backgrounds:}~Veto events with any OSSF dilepton pair satisfying $|m_{2\ell} - m_Z| < 5\;\GeV$. Also, since the signal arises from $Z\rightarrow 4\ell+X$ decays, veto events with any $m_{4\ell}+5\,\,\mathrm{GeV}> m_Z$.
  \item \emph{Suppression of combinatorics:}~When re-constructing $A'$ candidates among possible OSSF dilepton pairs, $A'$ candidates are chosen such that the dilepton pairs minimize $|m_i - m_j|$, where $i$ and $j$ refer to the $A'$ candidates and $m_i$ and $m_j$ refer to their respective masses.
\end{itemize}
We find the efficiency of these selections 
is $\approx 20-50\%$ for signal events (depending on the masses $\mAp$ and $\mHd$) and about $\approx 1\%$ for background.

Additionally, we exploit the fact that the four leptons in $\hd\rightarrow4\ell$ decays reconstruct $\mHd$.
A typical 4-lepton invariant mass resolution is approximately \cite{Chatrchyan:2012cg}
\beq\label{eq:resolution}
\Delta m_{4\ell} = 0.13\;\GeV + 0.065\mHd.
\eeq
Therefore, for each signal hypothesis mass $\mHd$, we define a bin centered at mass $\mHd$
 and width given by
Eq.~\eqref{eq:resolution}, and consider only events inside this bin.
 This eliminates much of the remaining background. We 
comment that all of our selection criteria are approximate and should be
re-optimized by the experimental collaborations once more accurate, data-driven background
 estimates are obtained.\\

\noindent {\bf Backgrounds:}~The dominant SM background by far is
the $\gamma/Z$-initiated inclusive $pp\rightarrow 4 \ell+X$ final
state.\footnote{This was also found to be the dominant background in both CMS
  low--$\slashed{E}_{\rm T}$ 4-lepton searches (with either $<2$ or $=2$ OSSF pairs)~\cite{Sirunyan:2017lae}.}
  Other backgrounds, such as $t\bar{t}$ and
  multi-boson production,  are subdominant after the cuts we apply and could be 
  further suppressed by additional requirements on $b$-jets, the maximum allowed
  lepton $\pT$, and the maximum $\slashed{E}_{\rm T}$.\\
  
\begin{figure*}
  \centering
  \includegraphics[width=0.8\textwidth]{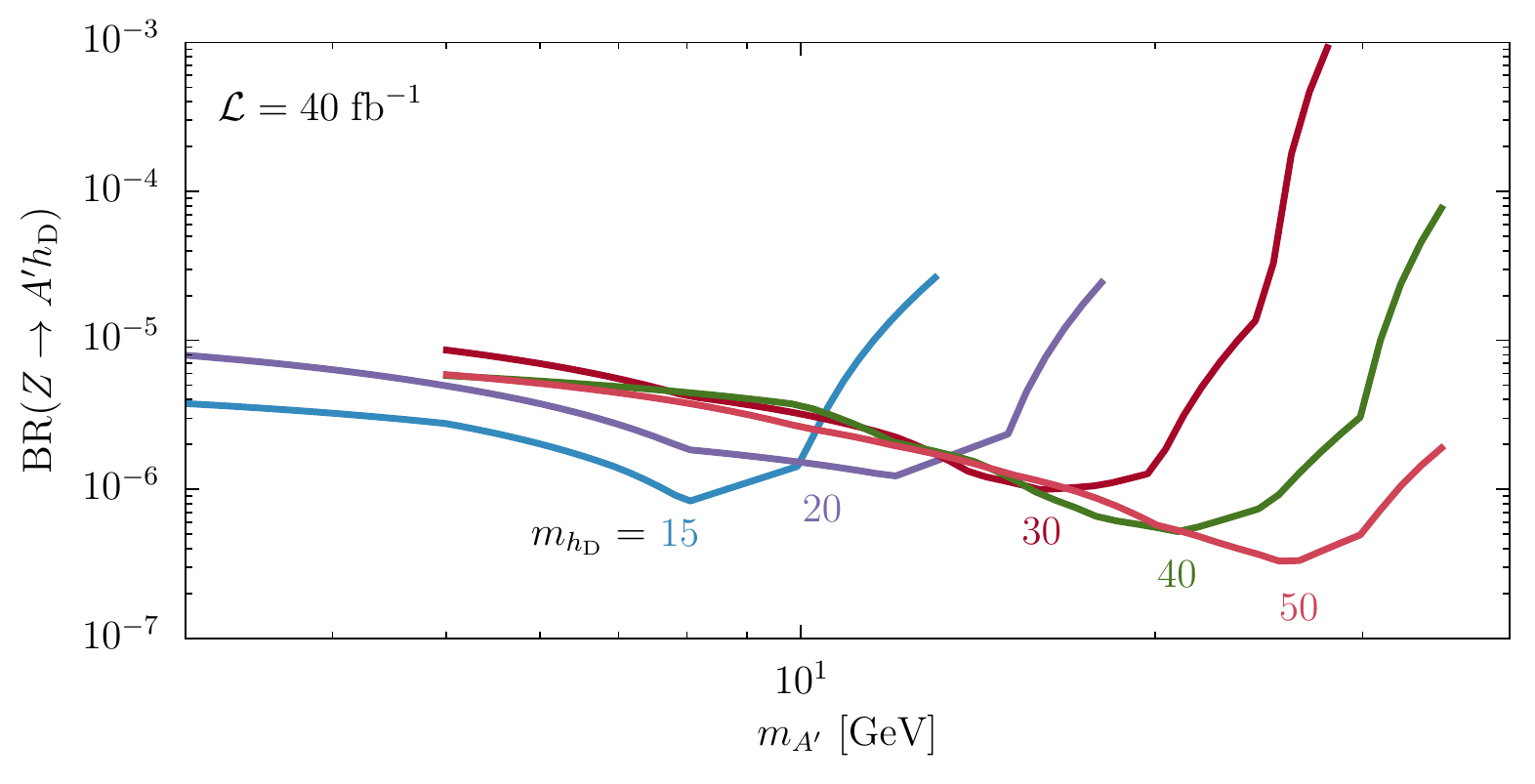}
  \caption{Projected $95\%$ c.l. sensitivity of an LHC search with $\mathcal{L} = 40\;\fb^{-1}$ at $\sqrt{s}=13$ TeV for $Z\rightarrow \hd \Ap \rightarrow 4\ell + X$ where $4\ell$ are required to reconstruct $\mHd$. The sensitivity
  is expressed in terms of the accessible branching fraction of $Z\rightarrow \hd \Ap$ decays.
  \label{fig:combined_sensitivity_br_to_ah_present}}
\end{figure*}

\begin{figure*}
  \centering
  \includegraphics[width=0.8\textwidth]{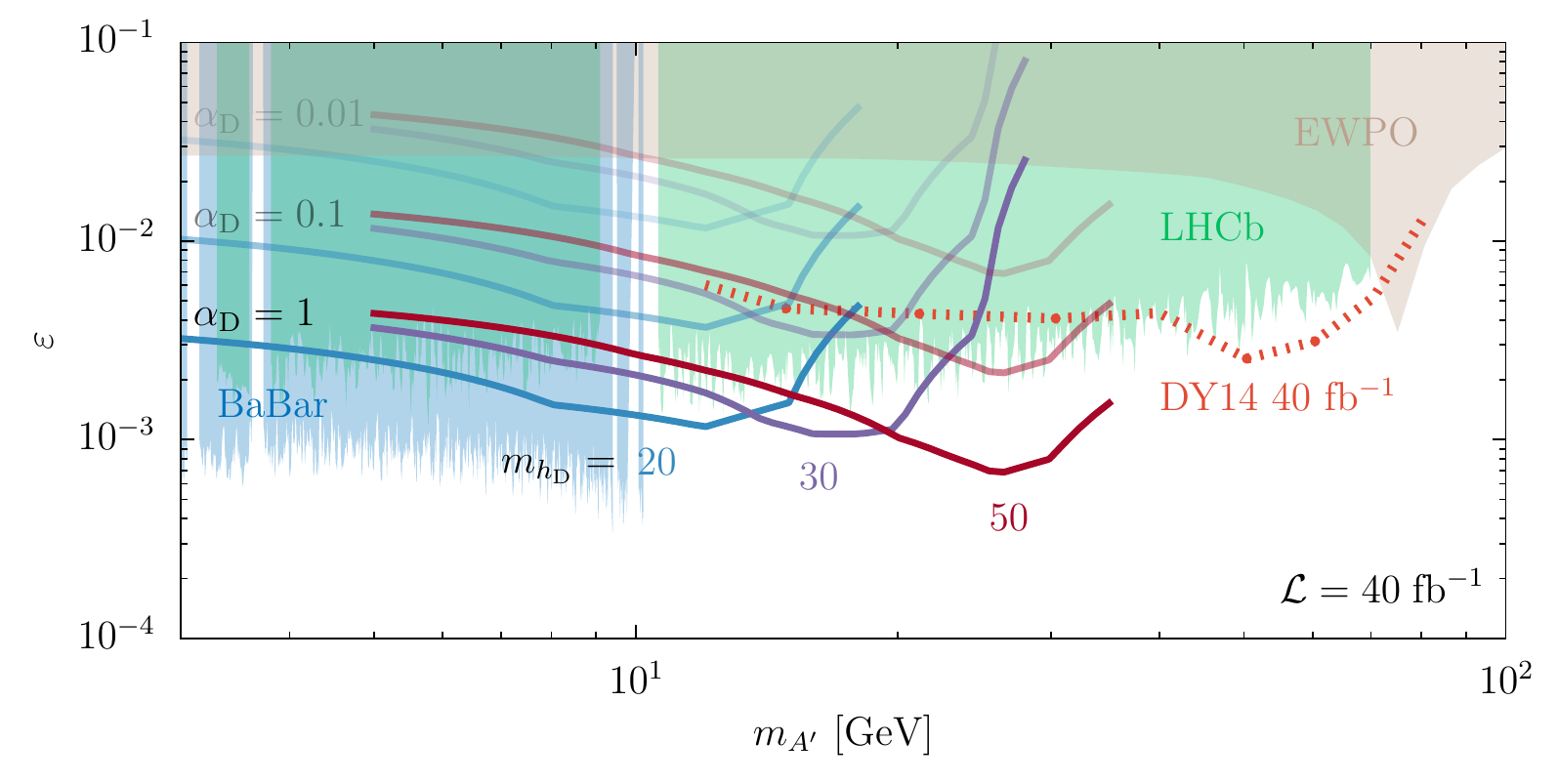}
  \caption{Projected $95\%$ c.l. sensitivity of an LHC search with $\mathcal{L}
  = 40\;\fb^{-1}$ at $\sqrt{s}=13$ TeV for $Z\rightarrow \hd \Ap \rightarrow
  4\ell + X$ where $4\ell$ are required to reconstruct $\mHd$. Each curve is
  labeled by the value of $\mHd$ in GeV. The dotted line gives the sensitivity
  of the Drell-Yan search from Ref.~\cite{Curtin:2014cca}. We also show
  existing constraints from electroweak precision
  observables~\cite{Curtin:2014cca}, BaBar~\cite{Lees:2014xha} and
  LHCb~\cite{Aaij:2017rft}.
The three sets of lines from bottom (dark) to top (light) correspond to $\ad = 1,\,0.1$, and $0.01$.
  \label{fig:prompt_sensitivity_alphaD}}
\end{figure*}

\begin{figure*}
  \centering
  \includegraphics[width=0.8\textwidth]{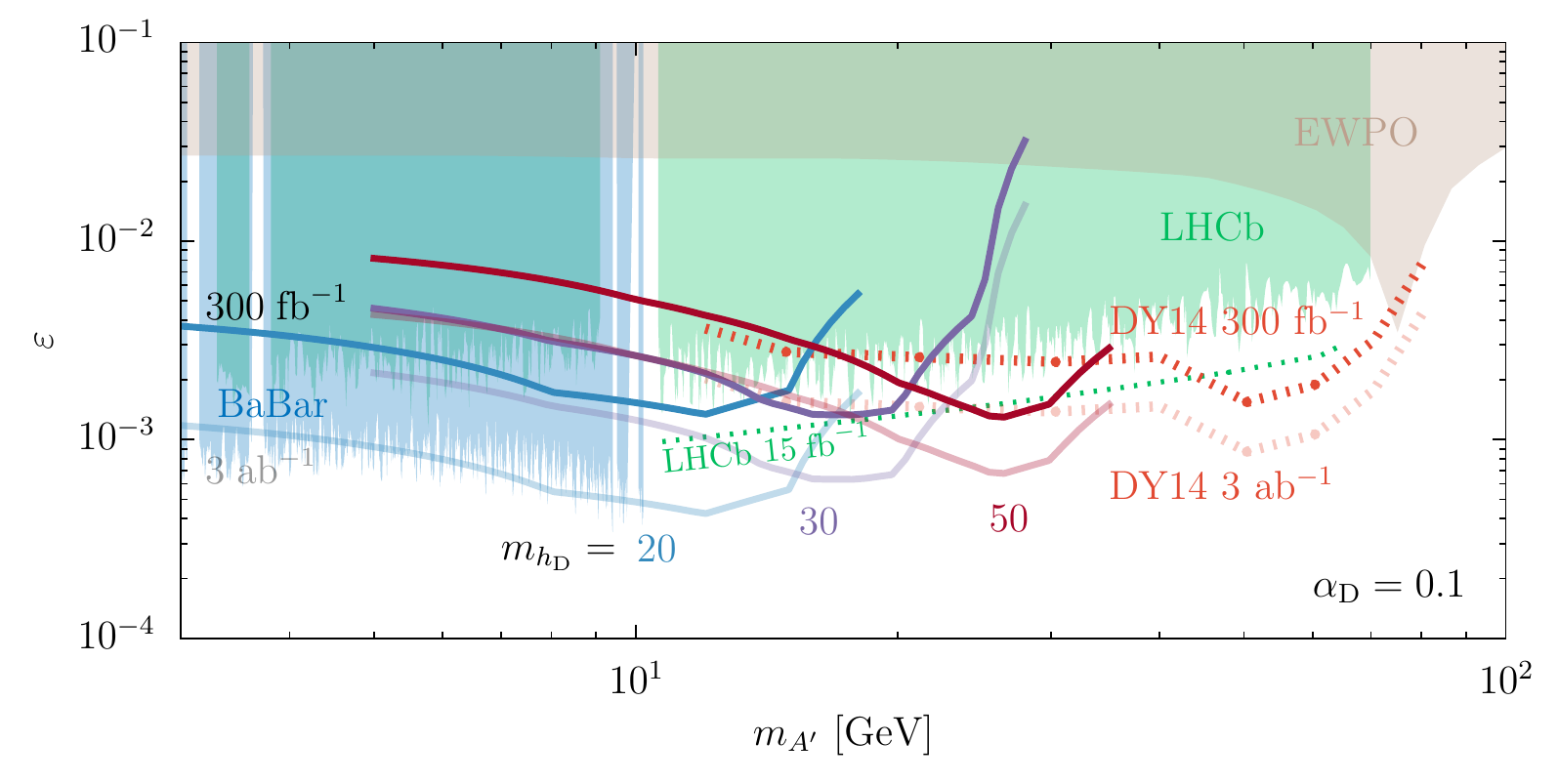}
  \caption{Sensitivity projections for the $Z\rightarrow \hd \Ap \rightarrow 4\ell + X$ search for integrated luminosity of $300\;\fb^{-1}$ 
  (upper dark lines) and $3000\;\fb^{-1}$ (lower faint lines) at $\sqrt{s}=13$ TeV and $\ad=0.1$. The dotted green line shows the projected 
  LHCb sensitivity with 15 $\fb^{-1}$ from Ref.~\cite{Ilten:2016tkc}, while the dashed lines show the projections
  for the proposed Drell-Yan search from Ref.~\cite{Curtin:2014cca}.
  Notation and existing bounds are the same as in Fig.~\ref{fig:prompt_sensitivity_alphaD}.
  \label{fig:prompt_sensitivity_future}}
\end{figure*}

\noindent {\bf Sensitivity Projections:}~We now make projections for the
sensitivity of the above proposed search to the dark Abelian Higgs model.
In addition to the selections described above, we apply a flat event-level $50\%$ penalty
for reconstructing the four soft leptons.
We evaluate the maximum number of allowed signal events at $95\%$ c.l.  assuming Poisson
statistics and a background-only hypothesis, and extract an estimated limit on the signal rate.
We first express the expected sensitivity in terms of the $Z\rightarrow A'h_{\rm D}$ branching
fraction, which we show for
$40\,\,\mathrm{fb}^{-1}$ in Fig.~\ref{fig:combined_sensitivity_br_to_ah_present}. 
This sensitivity depends only on the $A'$ and $\hd$ masses and not separately
on the hidden-sector couplings.
In Fig.~\ref{fig:prompt_sensitivity_alphaD}, we show the estimated sensitivity to
the kinetic mixing parameter, $\varepsilon$,
 for various values of $\ad$. Looking forward, in
Fig.~\ref{fig:prompt_sensitivity_future} we demonstrate for the
particular case $\ad=0.1$ the sensitivity that
can be achieved if the same trigger and analysis selections can be
maintained throughout Run 3 and the High-Luminosity LHC integrated luminosity
benchmarks.  For comparison with our results, we display the projected sensitivity from dilepton
resonance searches.

 We emphasize
that beyond allowing for a discovery of $\Ap$ and $\hd$, such a search can yield additional
information about the dark sector. In particular,
the $Z$ branching ratio into $\Ap\hd$ is sensitive to the dark coupling, $\ad$, which is readily 
seen in Fig.~\ref{fig:prompt_sensitivity_alphaD}. Thus, a combined discovery of $A'$ in Drell-Yan production
direct production and via $Z\rightarrow A'\hd$ would allow for the determination of
the hidden-sector coupling, $\ad$. The
$Z\rightarrow \hd A'$ branching fraction in Eq.~\eqref{eq:z_branching} depends
linearly on $\ad$ and quadratically on $\varepsilon$, and so the reach in $\varepsilon$ for other
values of $\ad$ can be obtained by rescaling our limits on $\varepsilon$ by $1/\sqrt{\ad}$.

\section{Displaced Dark Higgs Decays}\label{sec:displaced}

The results of Sec.~\ref{sec:prompt} focused exclusively on the case where
$\hd$ can  decay into at least one on-shell $A'$. In this
section we study the opposite mass hierarchy:~$\mHd < \mAp$.
In this regime, the decays $Z\rightarrow\Ap\hd$ are still allowed for $m_Z>m_{\Ap}+m_{\hd}$, but the $\hd$ can
only decay radiatively or via entirely off-shell $\Ap$. Consequently, the $\hd$ decay is typically displaced from
the primary interaction point. 
These signatures benefit from smaller
backgrounds compared to the prompt final states described in Section~\ref{sec:prompt}. We now
discuss the salient phenomenological features.

\begin{figure*}
  \centering
  \includegraphics[width=0.8\textwidth]{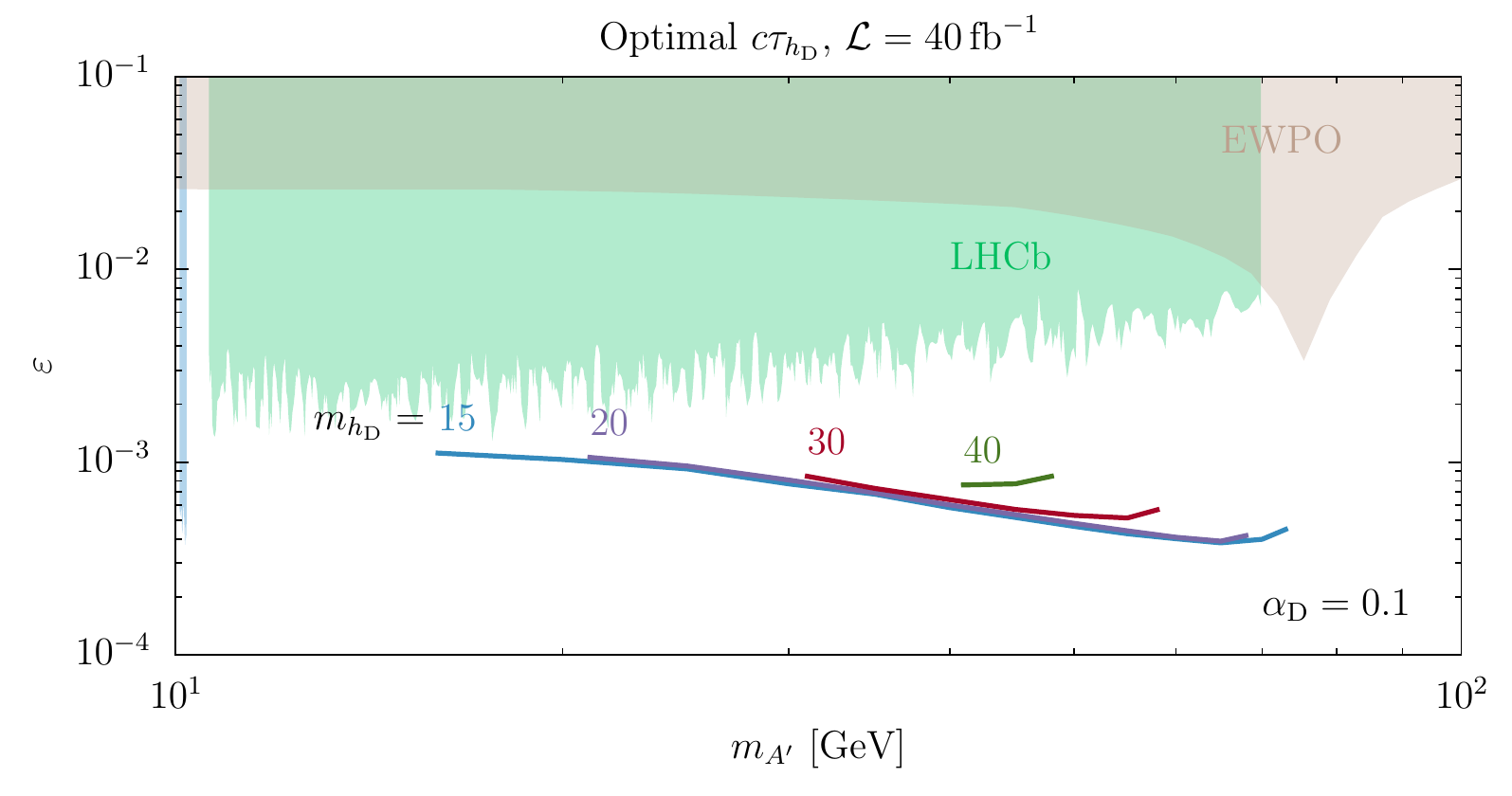}\\
  \includegraphics[width=0.8\textwidth]{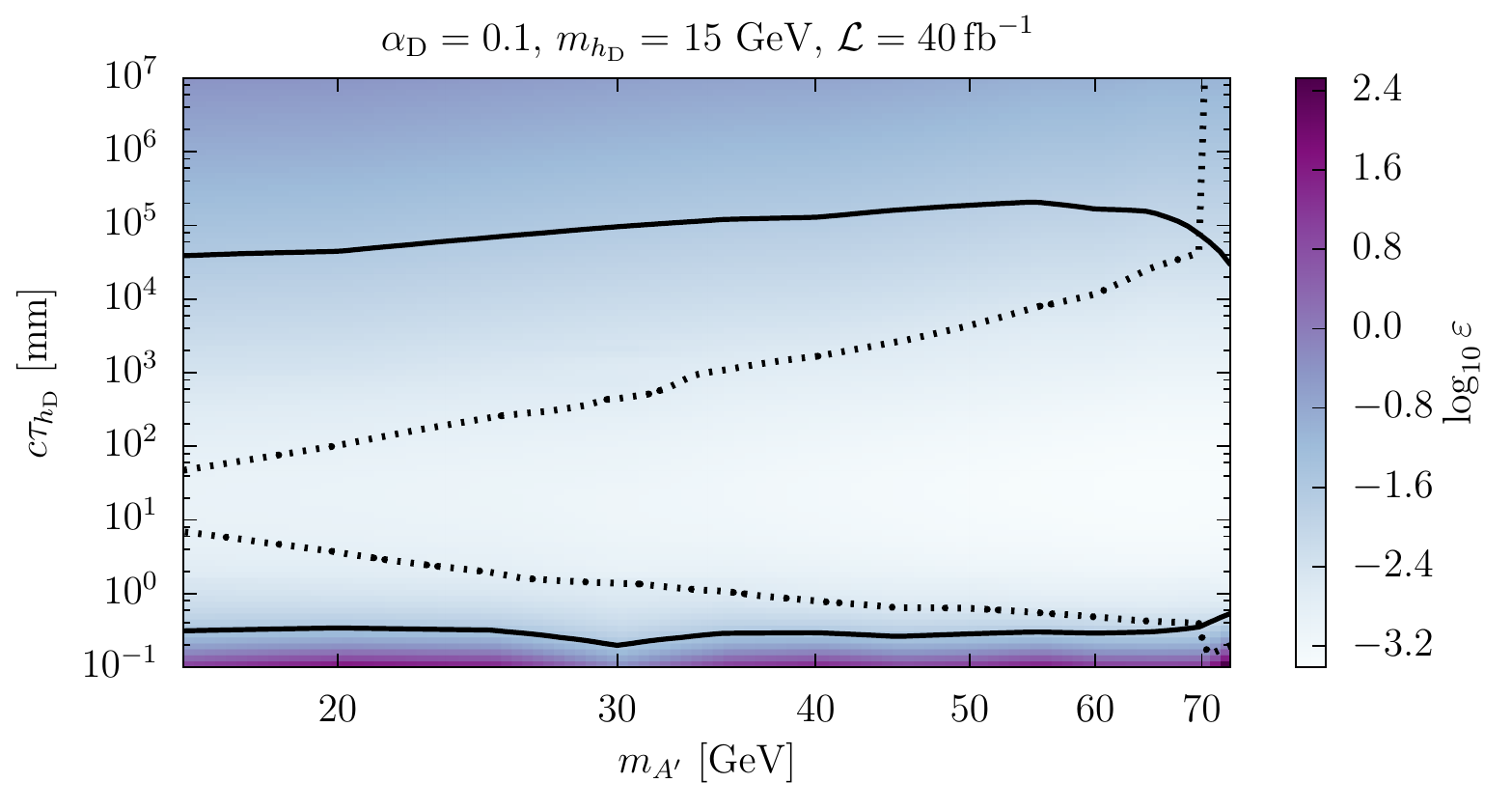}\\ 
  \caption{Projected sensitivity to the scenario where  $\mHd < \mAp$, which
    gives rise to displaced decays of the dark Higgs. Projections are shown for $\sqrt{s}=13$ TeV
    and $\mathcal{L} = 40\;\fb^{-1}$. In the top panel, we compute the
   sensitivity for 10 signal events after cuts to $\varepsilon$ in the $(\mAp, \varepsilon)$ plane for different 
    values of $\mHd$ (labeled in GeV); for each $\mAp$ point we have selected 
    the $\hd$ lifetime that gives the optimal reach in
    $\varepsilon$. In the bottom panel we 
    compute the expected sensitivity in the $(\mAp, c\tau_{\hd})$ plane for $\mHd = 15\;\GeV$. The
    projected reach of the displaced search exceeds the existing constraints
    from electroweak precision observables (LHCb prompt search, showing the envelope of the exclusion contour) within the solid (dotted) black contours.
  \label{fig:higgstrahlung_sensitivity_displaced}}
\end{figure*}

In the parameter space $m_{\hd} < m_{\Ap}$, the $\hd$ can still decay
into a two-body final state. For one possibility, $\hd$ decays through a  loop of virtual 
$\Ap$ into SM fermions. This partial width scales approximately as \cite{Batell:2009yf}
\be
\Gamma(\hd\rightarrow f \bar{f}) & \sim & \frac{\alpha^2 Q_f^4 \ad \varepsilon^4}{32\pi^2}\left(\frac{m_f}{\mAp}\right)^2 \mHd\label{eq:width-displaced-hypercharge}\\
&\sim& \left(1\,\,\mathrm{m}^{-1}\right)\left(\frac{\ad}{0.1}\right)\left(\frac{\varepsilon}{10^{-2}}\right)^4\left(\frac{15\,\,\mathrm{GeV}}{\mAp}\right)\nonumber
\ee
when summing over SM final-state fermions and taking $\mAp\sim \mHd$. 
At this order of $\varepsilon^4$, there is also a tree-level four-body decay of the
$\hd$ via two off-shell $\Ap$. However, for the moderate dark Higgs
masses that we are interested in, the loop decay typically dominates. 

So far, we have assumed that the mixing between the $\hd$ and the SM Higgs
is zero, and consequently the
only allowed $\hd$ decay modes are via its coupling with $\Ap$.
However, a mixing between the $\hd$ and the SM Higgs is {\it induced} 
 by a loop of dark vectors. This mixing is
given by
\be
V(H, \Hd) &\supset& \kappa(\mu) \left | H \right | ^2 \left | \Hd \right | ^2,
\label{eq:induced_mixing_potential}
\ee
where the renormalization-scale-dependent mixing term scales as
\be\label{eq:induced_kappa}
\kappa(\mu) &\sim& \frac{\alpha\, \ad\,\varepsilon^2}{(4\pi)^2} \log\left(\frac{\mu}{\Lambda}\right)+\kappa(\Lambda),
\ee
where $\Lambda$ is an ultraviolet (UV) energy scale. The decay $\hd\rightarrow f\bar{f}$ due to the the mixing in Eq.~\eqref{eq:induced_kappa}
 is parametrically similar to the $\Ap$-loop-induced and four-body decay modes. 
 The precise value of this loop-induced mixing depends explicitly on the UV 
 value of $\kappa$. The mixing $\kappa$ can in principle be zero in the infrared,
but this  represents a tuning of model parameters.
Thus, the decay width of $\hd$ for $\mAp > \mHd$ (and whether it proceeds radiatively
or via Higgs mixing) depends sensitively on 
the value of $\kappa$ in the UV and its renormalization-group evolution.

Because of this model dependence,  we take a bottom-up
approach and focus on the plausible signatures of $\hd$ decay for $m_{\hd}<m_{\Ap}$,
 all of which  feature
decays of $\hd$ at a displaced vertex for the parameter space that is accessible to
 the LHC. The dominant displaced signatures that can arise are:
\begin{itemize}
\item Displaced decay of $\hd$ into SM fermions according to Eq.~\eqref{eq:width-displaced-hypercharge}.
The branching fractions into heavy-flavor objects ($b$, $c$, and $\tau$) are comparable:~the color factors and/or
heavier masses of the $b/c$ are compensated by the smaller electric charges.
This case will arise when the branching ratio given by
Eq.~\eqref{eq:width-displaced-hypercharge} dominates. The $\hd$
will then give (at least) two displaced tracks, which can be leptons or hadrons. The
combined final state from the rare $Z$ decay is two prompt leptons in
association with the two or more displaced tracks.  
\item Displaced decay of $\hd$ through Higgs mixing in Eq.~\eqref{eq:induced_mixing_potential}. 
This decay occurs predominantly into bottom quarks. We require displaced tracks
from the $b$-quark hadronization but apply no $b$-tagging requirements. The displaced
vertex comes in
  association with prompt leptons from the $A'$ decay in $Z\rightarrow A'\hd\rightarrow \ell^+\ell^- \hd$.
\item The sub-dominant decay of $\hd$ into four SM fermions can give a rather striking final state; we do not
consider it further, although it could give rise to an interesting signature for future study.
\end{itemize}

We estimate the sensitivity for a prompt, two-lepton final state in
association with displaced tracks (leptons or hadrons) originating
from a displaced vertex as follows.
The signal events are selected using standard 
dilepton triggers~\cite{Sirunyan:2017lae}
\begin{itemize}
  \item two OSSF muons with $\pT > 17,\; 8\;$ GeV, or
  \item two OSSF electrons with $\pT > 23,\; 12$ GeV.
\end{itemize}
We further require the track transverse impact parameters, $|d_0|$, to 
lie within $ 1\;\mm< |d_0| < 200\;\mm$ (motivated collectively by the
 ATLAS~\cite{Aad:2015rba} and CMS~\cite{CMS:2014hka} $|d_0|$ reconstruction capability).
 We further require the point of $\hd$ decay to  occur within 
$200$ mm of the primary vertex in both the transverse and longitudinal directions. Finally, 
we apply an efficiency for selection of a displaced vertex. This efficiency depends on the details
of the experimental search, and in particular the need to reject certain backgrounds. In existing searches,
the efficiencies for displaced vertices in the inner detector vary widely from $\sim10-30\%$ \cite{Aad:2015rba}   through to $\sim50\%$ \cite{CMS:2014hka}. The signal also features a resonant $A'$ mass that can be 
reconstructed in the prompt leptons:~therefore, backgrounds are  lower than for inclusive displaced vertex
searches, and data-driven background estimation is more straightforward in the variable $m_{\ell^+\ell^-}$ from the prompt leptons. Therefore, we prioritize signal efficiency and choose to apply a flat 50\% vertex tagging efficiency.
The sensitivity of this search is estimated by requiring an observation of 10 signal events, assuming no background.

In Fig.~\ref{fig:higgstrahlung_sensitivity_displaced} we show the projected
sensitivity in two ways. In the top panel we compute the sensitivity to $\varepsilon$
in the $(\mAp, \varepsilon)$ plane, where for each choice of masses we have selected the $\hd$
lifetime that gives the optimal reach in $\varepsilon$. In the bottom panel we select a
specific value of $\mHd$ and show the expected sensitivity in the
$(\mAp,c\tau_{\hd})$ plane. The projected sensitivity of this search exceeds the 
current limits from EWPO (LHCb) inside the solid black (dotted) contours.
With current
levels of data, one can already probe $\varepsilon$ at or better than the level of $10^{-3}$.

\section{Discussion and Conclusions}\label{sec:discussion}

In this article, we have shown that rare decays of the Standard Model $Z$ boson are
powerful probes of hidden sectors. Using a dark Abelian Higgs model as a benchmark, we have
demonstrated how studying rare $Z$ decays into hidden sector particles can allow for the discovery of exotic particles 
through the same interactions responsible for generating hidden-sector particle masses. 

 $Z$ decays into  hidden-sector particles typically give rise to large multiplicities of soft
particles. When the hidden-sector particles can decay leptonically, such as in the dark Abelian Higgs scenario, this results in striking events with up to six soft leptons. 
We have demonstrated a range of prompt and displaced signatures that can occur in this
model, providing a path for experimentally
discovering or constraining these new particles.

Crucial to the success of the strategies we outline in this article is the fact that trigger and reconstruction
thresholds for leptons must be kept low. Because the $Z$ boson mass-energy is distributed among
six or more particles, a moderate increase in trigger thresholds would greatly curtail the sensitivity of
ATLAS or CMS to high-multiplicity hidden-sector signatures. While the high-luminosity running of the LHC
will introduce new challenges such as increased pile-up, we urge the experimental collaborations to consider
maintaining low-threshold, high-multiplicity triggers where possible 
(including potentially triggers requiring four or more leptons) in
order to retain sensitivity to well-motivated hidden-sector models. Furthermore, our study has
focused on only one particular high-multiplicity hidden-sector signature:~models giving rise to even higher multiplicities are
possible, motivating keeping high-multiplicity trigger thresholds as low as possible.

The searches we propose are complementary to existing proposals for observation
of the direct production of the dark photon, $A'$. A discovery in both channels
could allow for the determination of the $A'$ mixing with the SM, the 
hidden-sector gauge coupling, and the masses of the dark photon and dark Higgs. 
A comprehensive characterization of the properties of the hidden sector is 
therefore possible at the LHC.

Finally, we have focused on the specific case where the $Z$ boson can decay  to
a dark gauge boson and a dark scalar where both are on-shell. There exist
spectra where this is not the case, such as when $m_Z<\mAp+\mHd$, and there may
yet be other interesting signatures of dark Higgs-strahlung originating from
off-shell $Z$ production. Indeed, the signatures we discuss in this article
should be  applicable to more general mass hierarchies, and we encourage ATLAS,
CMS, and LHCb to avoid overly optimizing their search strategies to the
particular examples studied here where possible.

\acknowledgments
We would like to thank Justin Chiu, Michel Lefebvre, Michael Peskin and Maxim Pospelov for helpful conversations. EI is
supported by the United States Department of Energy under
Grant Contract desc0012704. NB is supported by the United States Department of Energy under Contract No. DE-AC02-76SF00515.
We express our gratitude to the Aspen Center for Physics, where part of this work was completed and which is supported by National Science Foundation grant PHY-1607761.

\bibliography{biblio}

\begin{thebibliography}{59}%
\makeatletter
\providecommand \@ifxundefined [1]{%
 \@ifx{#1\undefined}
}%
\providecommand \@ifnum [1]{%
 \ifnum #1\expandafter \@firstoftwo
 \else \expandafter \@secondoftwo
 \fi
}%
\providecommand \@ifx [1]{%
 \ifx #1\expandafter \@firstoftwo
 \else \expandafter \@secondoftwo
 \fi
}%
\providecommand \natexlab [1]{#1}%
\providecommand \enquote  [1]{``#1''}%
\providecommand \bibnamefont  [1]{#1}%
\providecommand \bibfnamefont [1]{#1}%
\providecommand \citenamefont [1]{#1}%
\providecommand \href@noop [0]{\@secondoftwo}%
\providecommand \href [0]{\begingroup \@sanitize@url \@href}%
\providecommand \@href[1]{\@@startlink{#1}\@@href}%
\providecommand \@@href[1]{\endgroup#1\@@endlink}%
\providecommand \@sanitize@url [0]{\catcode `\\12\catcode `\$12\catcode
  `\&12\catcode `\#12\catcode `\^12\catcode `\_12\catcode `\%12\relax}%
\providecommand \@@startlink[1]{}%
\providecommand \@@endlink[0]{}%
\providecommand \url  [0]{\begingroup\@sanitize@url \@url }%
\providecommand \@url [1]{\endgroup\@href {#1}{\urlprefix }}%
\providecommand \urlprefix  [0]{URL }%
\providecommand \Eprint [0]{\href }%
\providecommand \doibase [0]{http://dx.doi.org/}%
\providecommand \selectlanguage [0]{\@gobble}%
\providecommand \bibinfo  [0]{\@secondoftwo}%
\providecommand \bibfield  [0]{\@secondoftwo}%
\providecommand \translation [1]{[#1]}%
\providecommand \BibitemOpen [0]{}%
\providecommand \bibitemStop [0]{}%
\providecommand \bibitemNoStop [0]{.\EOS\space}%
\providecommand \EOS [0]{\spacefactor3000\relax}%
\providecommand \BibitemShut  [1]{\csname bibitem#1\endcsname}%
\let\auto@bib@innerbib\@empty
\bibitem [{\citenamefont {Chen}\ \emph {et~al.}(1996)\citenamefont {Chen},
  \citenamefont {Drees},\ and\ \citenamefont {Gunion}}]{Chen:1995yu}%
  \BibitemOpen
  \bibfield  {author} {\bibinfo {author} {\bibfnamefont {C.~H.}\ \bibnamefont
  {Chen}}, \bibinfo {author} {\bibfnamefont {M.}~\bibnamefont {Drees}}, \ and\
  \bibinfo {author} {\bibfnamefont {J.~F.}\ \bibnamefont {Gunion}},\ }\href
  {\doibase 10.1103/PhysRevLett.76.2002} {\bibfield  {journal} {\bibinfo
  {journal} {Phys. Rev. Lett.}\ }\textbf {\bibinfo {volume} {76}},\ \bibinfo
  {pages} {2002} (\bibinfo {year} {1996})},\ \Eprint
  {http://arxiv.org/abs/hep-ph/9512230} {arXiv:hep-ph/9512230 [hep-ph]}
  \BibitemShut {NoStop}%
\bibitem [{\citenamefont {Thomas}\ and\ \citenamefont
  {Wells}(1998)}]{Thomas:1998wy}%
  \BibitemOpen
  \bibfield  {author} {\bibinfo {author} {\bibfnamefont {S.~D.}\ \bibnamefont
  {Thomas}}\ and\ \bibinfo {author} {\bibfnamefont {J.~D.}\ \bibnamefont
  {Wells}},\ }\href {\doibase 10.1103/PhysRevLett.81.34} {\bibfield  {journal}
  {\bibinfo  {journal} {Phys. Rev. Lett.}\ }\textbf {\bibinfo {volume} {81}},\
  \bibinfo {pages} {34} (\bibinfo {year} {1998})},\ \Eprint
  {http://arxiv.org/abs/hep-ph/9804359} {arXiv:hep-ph/9804359 [hep-ph]}
  \BibitemShut {NoStop}%
\bibitem [{\citenamefont {Han}\ \emph {et~al.}(2014)\citenamefont {Han},
  \citenamefont {Kobakhidze}, \citenamefont {Liu}, \citenamefont {Saavedra},
  \citenamefont {Wu},\ and\ \citenamefont {Yang}}]{Han:2013usa}%
  \BibitemOpen
  \bibfield  {author} {\bibinfo {author} {\bibfnamefont {C.}~\bibnamefont
  {Han}}, \bibinfo {author} {\bibfnamefont {A.}~\bibnamefont {Kobakhidze}},
  \bibinfo {author} {\bibfnamefont {N.}~\bibnamefont {Liu}}, \bibinfo {author}
  {\bibfnamefont {A.}~\bibnamefont {Saavedra}}, \bibinfo {author}
  {\bibfnamefont {L.}~\bibnamefont {Wu}}, \ and\ \bibinfo {author}
  {\bibfnamefont {J.~M.}\ \bibnamefont {Yang}},\ }\href {\doibase
  10.1007/JHEP02(2014)049} {\bibfield  {journal} {\bibinfo  {journal} {JHEP}\
  }\textbf {\bibinfo {volume} {02}},\ \bibinfo {pages} {049} (\bibinfo {year}
  {2014})},\ \Eprint {http://arxiv.org/abs/1310.4274} {arXiv:1310.4274
  [hep-ph]} \BibitemShut {NoStop}%
\bibitem [{\citenamefont {Low}\ and\ \citenamefont {Wang}(2014)}]{Low:2014cba}%
  \BibitemOpen
  \bibfield  {author} {\bibinfo {author} {\bibfnamefont {M.}~\bibnamefont
  {Low}}\ and\ \bibinfo {author} {\bibfnamefont {L.-T.}\ \bibnamefont {Wang}},\
  }\href {\doibase 10.1007/JHEP08(2014)161} {\bibfield  {journal} {\bibinfo
  {journal} {JHEP}\ }\textbf {\bibinfo {volume} {08}},\ \bibinfo {pages} {161}
  (\bibinfo {year} {2014})},\ \Eprint {http://arxiv.org/abs/1404.0682}
  {arXiv:1404.0682 [hep-ph]} \BibitemShut {NoStop}%
\bibitem [{\citenamefont {Alwall}\ \emph
  {et~al.}(2009{\natexlab{a}})\citenamefont {Alwall}, \citenamefont {Le},
  \citenamefont {Lisanti},\ and\ \citenamefont {Wacker}}]{Alwall:2008va}%
  \BibitemOpen
  \bibfield  {author} {\bibinfo {author} {\bibfnamefont {J.}~\bibnamefont
  {Alwall}}, \bibinfo {author} {\bibfnamefont {M.-P.}\ \bibnamefont {Le}},
  \bibinfo {author} {\bibfnamefont {M.}~\bibnamefont {Lisanti}}, \ and\
  \bibinfo {author} {\bibfnamefont {J.~G.}\ \bibnamefont {Wacker}},\ }\href
  {\doibase 10.1103/PhysRevD.79.015005} {\bibfield  {journal} {\bibinfo
  {journal} {Phys. Rev.}\ }\textbf {\bibinfo {volume} {D79}},\ \bibinfo {pages}
  {015005} (\bibinfo {year} {2009}{\natexlab{a}})},\ \Eprint
  {http://arxiv.org/abs/0809.3264} {arXiv:0809.3264 [hep-ph]} \BibitemShut
  {NoStop}%
\bibitem [{\citenamefont {Fan}\ \emph {et~al.}(2011)\citenamefont {Fan},
  \citenamefont {Reece},\ and\ \citenamefont {Ruderman}}]{Fan:2011yu}%
  \BibitemOpen
  \bibfield  {author} {\bibinfo {author} {\bibfnamefont {J.}~\bibnamefont
  {Fan}}, \bibinfo {author} {\bibfnamefont {M.}~\bibnamefont {Reece}}, \ and\
  \bibinfo {author} {\bibfnamefont {J.~T.}\ \bibnamefont {Ruderman}},\ }\href
  {\doibase 10.1007/JHEP11(2011)012} {\bibfield  {journal} {\bibinfo  {journal}
  {JHEP}\ }\textbf {\bibinfo {volume} {11}},\ \bibinfo {pages} {012} (\bibinfo
  {year} {2011})},\ \Eprint {http://arxiv.org/abs/1105.5135} {arXiv:1105.5135
  [hep-ph]} \BibitemShut {NoStop}%
\bibitem [{\citenamefont {Evans}\ and\ \citenamefont
  {Kats}(2013)}]{Evans:2012bf}%
  \BibitemOpen
  \bibfield  {author} {\bibinfo {author} {\bibfnamefont {J.~A.}\ \bibnamefont
  {Evans}}\ and\ \bibinfo {author} {\bibfnamefont {Y.}~\bibnamefont {Kats}},\
  }\href {\doibase 10.1007/JHEP04(2013)028} {\bibfield  {journal} {\bibinfo
  {journal} {JHEP}\ }\textbf {\bibinfo {volume} {04}},\ \bibinfo {pages} {028}
  (\bibinfo {year} {2013})},\ \Eprint {http://arxiv.org/abs/1209.0764}
  {arXiv:1209.0764 [hep-ph]} \BibitemShut {NoStop}%
\bibitem [{\citenamefont {Boehm}\ and\ \citenamefont
  {Fayet}(2004)}]{Boehm:2003hm}%
  \BibitemOpen
  \bibfield  {author} {\bibinfo {author} {\bibfnamefont {C.}~\bibnamefont
  {Boehm}}\ and\ \bibinfo {author} {\bibfnamefont {P.}~\bibnamefont {Fayet}},\
  }\href {\doibase 10.1016/j.nuclphysb.2004.01.015} {\bibfield  {journal}
  {\bibinfo  {journal} {Nucl. Phys.}\ }\textbf {\bibinfo {volume} {B683}},\
  \bibinfo {pages} {219} (\bibinfo {year} {2004})},\ \Eprint
  {http://arxiv.org/abs/hep-ph/0305261} {arXiv:hep-ph/0305261 [hep-ph]}
  \BibitemShut {NoStop}%
\bibitem [{\citenamefont {Pospelov}(2009)}]{Pospelov:2008zw}%
  \BibitemOpen
  \bibfield  {author} {\bibinfo {author} {\bibfnamefont {M.}~\bibnamefont
  {Pospelov}},\ }\href {\doibase 10.1103/PhysRevD.80.095002} {\bibfield
  {journal} {\bibinfo  {journal} {Phys. Rev.}\ }\textbf {\bibinfo {volume}
  {D80}},\ \bibinfo {pages} {095002} (\bibinfo {year} {2009})},\ \Eprint
  {http://arxiv.org/abs/0811.1030} {arXiv:0811.1030 [hep-ph]} \BibitemShut
  {NoStop}%
\bibitem [{\citenamefont {Strassler}\ and\ \citenamefont
  {Zurek}(2007)}]{Strassler:2006im}%
  \BibitemOpen
  \bibfield  {author} {\bibinfo {author} {\bibfnamefont {M.~J.}\ \bibnamefont
  {Strassler}}\ and\ \bibinfo {author} {\bibfnamefont {K.~M.}\ \bibnamefont
  {Zurek}},\ }\href {\doibase 10.1016/j.physletb.2007.06.055} {\bibfield
  {journal} {\bibinfo  {journal} {Phys. Lett.}\ }\textbf {\bibinfo {volume}
  {B651}},\ \bibinfo {pages} {374} (\bibinfo {year} {2007})},\ \Eprint
  {http://arxiv.org/abs/hep-ph/0604261} {arXiv:hep-ph/0604261 [hep-ph]}
  \BibitemShut {NoStop}%
\bibitem [{\citenamefont {Strassler}\ and\ \citenamefont
  {Zurek}(2008)}]{Strassler:2006ri}%
  \BibitemOpen
  \bibfield  {author} {\bibinfo {author} {\bibfnamefont {M.~J.}\ \bibnamefont
  {Strassler}}\ and\ \bibinfo {author} {\bibfnamefont {K.~M.}\ \bibnamefont
  {Zurek}},\ }\href {\doibase 10.1016/j.physletb.2008.02.008} {\bibfield
  {journal} {\bibinfo  {journal} {Phys. Lett.}\ }\textbf {\bibinfo {volume}
  {B661}},\ \bibinfo {pages} {263} (\bibinfo {year} {2008})},\ \Eprint
  {http://arxiv.org/abs/hep-ph/0605193} {arXiv:hep-ph/0605193 [hep-ph]}
  \BibitemShut {NoStop}%
\bibitem [{\citenamefont {Han}\ \emph {et~al.}(2008)\citenamefont {Han},
  \citenamefont {Si}, \citenamefont {Zurek},\ and\ \citenamefont
  {Strassler}}]{Han:2007ae}%
  \BibitemOpen
  \bibfield  {author} {\bibinfo {author} {\bibfnamefont {T.}~\bibnamefont
  {Han}}, \bibinfo {author} {\bibfnamefont {Z.}~\bibnamefont {Si}}, \bibinfo
  {author} {\bibfnamefont {K.~M.}\ \bibnamefont {Zurek}}, \ and\ \bibinfo
  {author} {\bibfnamefont {M.~J.}\ \bibnamefont {Strassler}},\ }\href {\doibase
  10.1088/1126-6708/2008/07/008} {\bibfield  {journal} {\bibinfo  {journal}
  {JHEP}\ }\textbf {\bibinfo {volume} {07}},\ \bibinfo {pages} {008} (\bibinfo
  {year} {2008})},\ \Eprint {http://arxiv.org/abs/0712.2041} {arXiv:0712.2041
  [hep-ph]} \BibitemShut {NoStop}%
\bibitem [{\citenamefont {Batell}\ \emph {et~al.}(2009)\citenamefont {Batell},
  \citenamefont {Pospelov},\ and\ \citenamefont {Ritz}}]{Batell:2009yf}%
  \BibitemOpen
  \bibfield  {author} {\bibinfo {author} {\bibfnamefont {B.}~\bibnamefont
  {Batell}}, \bibinfo {author} {\bibfnamefont {M.}~\bibnamefont {Pospelov}}, \
  and\ \bibinfo {author} {\bibfnamefont {A.}~\bibnamefont {Ritz}},\ }\href
  {\doibase 10.1103/PhysRevD.79.115008} {\bibfield  {journal} {\bibinfo
  {journal} {Phys. Rev.}\ }\textbf {\bibinfo {volume} {D79}},\ \bibinfo {pages}
  {115008} (\bibinfo {year} {2009})},\ \Eprint {http://arxiv.org/abs/0903.0363}
  {arXiv:0903.0363 [hep-ph]} \BibitemShut {NoStop}%
\bibitem [{\citenamefont {Batell}\ \emph {et~al.}(2011)\citenamefont {Batell},
  \citenamefont {Pospelov},\ and\ \citenamefont {Ritz}}]{Batell:2009jf}%
  \BibitemOpen
  \bibfield  {author} {\bibinfo {author} {\bibfnamefont {B.}~\bibnamefont
  {Batell}}, \bibinfo {author} {\bibfnamefont {M.}~\bibnamefont {Pospelov}}, \
  and\ \bibinfo {author} {\bibfnamefont {A.}~\bibnamefont {Ritz}},\ }\href
  {\doibase 10.1103/PhysRevD.83.054005} {\bibfield  {journal} {\bibinfo
  {journal} {Phys. Rev.}\ }\textbf {\bibinfo {volume} {D83}},\ \bibinfo {pages}
  {054005} (\bibinfo {year} {2011})},\ \Eprint {http://arxiv.org/abs/0911.4938}
  {arXiv:0911.4938 [hep-ph]} \BibitemShut {NoStop}%
\bibitem [{\citenamefont {Arkani-Hamed}\ and\ \citenamefont
  {Weiner}(2008)}]{ArkaniHamed:2008qp}%
  \BibitemOpen
  \bibfield  {author} {\bibinfo {author} {\bibfnamefont {N.}~\bibnamefont
  {Arkani-Hamed}}\ and\ \bibinfo {author} {\bibfnamefont {N.}~\bibnamefont
  {Weiner}},\ }\href {\doibase 10.1088/1126-6708/2008/12/104} {\bibfield
  {journal} {\bibinfo  {journal} {JHEP}\ }\textbf {\bibinfo {volume} {12}},\
  \bibinfo {pages} {104} (\bibinfo {year} {2008})},\ \Eprint
  {http://arxiv.org/abs/0810.0714} {arXiv:0810.0714 [hep-ph]} \BibitemShut
  {NoStop}%
\bibitem [{\citenamefont {Cheung}\ \emph {et~al.}(2010)\citenamefont {Cheung},
  \citenamefont {Ruderman}, \citenamefont {Wang},\ and\ \citenamefont
  {Yavin}}]{Cheung:2009su}%
  \BibitemOpen
  \bibfield  {author} {\bibinfo {author} {\bibfnamefont {C.}~\bibnamefont
  {Cheung}}, \bibinfo {author} {\bibfnamefont {J.~T.}\ \bibnamefont
  {Ruderman}}, \bibinfo {author} {\bibfnamefont {L.-T.}\ \bibnamefont {Wang}},
  \ and\ \bibinfo {author} {\bibfnamefont {I.}~\bibnamefont {Yavin}},\ }\href
  {\doibase 10.1007/JHEP04(2010)116} {\bibfield  {journal} {\bibinfo  {journal}
  {JHEP}\ }\textbf {\bibinfo {volume} {04}},\ \bibinfo {pages} {116} (\bibinfo
  {year} {2010})},\ \Eprint {http://arxiv.org/abs/0909.0290} {arXiv:0909.0290
  [hep-ph]} \BibitemShut {NoStop}%
\bibitem [{\citenamefont {Baumgart}\ \emph {et~al.}(2009)\citenamefont
  {Baumgart}, \citenamefont {Cheung}, \citenamefont {Ruderman}, \citenamefont
  {Wang},\ and\ \citenamefont {Yavin}}]{Baumgart:2009tn}%
  \BibitemOpen
  \bibfield  {author} {\bibinfo {author} {\bibfnamefont {M.}~\bibnamefont
  {Baumgart}}, \bibinfo {author} {\bibfnamefont {C.}~\bibnamefont {Cheung}},
  \bibinfo {author} {\bibfnamefont {J.~T.}\ \bibnamefont {Ruderman}}, \bibinfo
  {author} {\bibfnamefont {L.-T.}\ \bibnamefont {Wang}}, \ and\ \bibinfo
  {author} {\bibfnamefont {I.}~\bibnamefont {Yavin}},\ }\href {\doibase
  10.1088/1126-6708/2009/04/014} {\bibfield  {journal} {\bibinfo  {journal}
  {JHEP}\ }\textbf {\bibinfo {volume} {04}},\ \bibinfo {pages} {014} (\bibinfo
  {year} {2009})},\ \Eprint {http://arxiv.org/abs/0901.0283} {arXiv:0901.0283
  [hep-ph]} \BibitemShut {NoStop}%
\bibitem [{\citenamefont {Curtin}\ \emph {et~al.}(2014)\citenamefont {Curtin}
  \emph {et~al.}}]{Curtin:2013fra}%
  \BibitemOpen
  \bibfield  {author} {\bibinfo {author} {\bibfnamefont {D.}~\bibnamefont
  {Curtin}} \emph {et~al.},\ }\href {\doibase 10.1103/PhysRevD.90.075004}
  {\bibfield  {journal} {\bibinfo  {journal} {Phys. Rev.}\ }\textbf {\bibinfo
  {volume} {D90}},\ \bibinfo {pages} {075004} (\bibinfo {year} {2014})},\
  \Eprint {http://arxiv.org/abs/1312.4992} {arXiv:1312.4992 [hep-ph]}
  \BibitemShut {NoStop}%
\bibitem [{\citenamefont {Aad}\ \emph {et~al.}(2014)\citenamefont {Aad} \emph
  {et~al.}}]{Aad:2014yea}%
  \BibitemOpen
  \bibfield  {author} {\bibinfo {author} {\bibfnamefont {G.}~\bibnamefont
  {Aad}} \emph {et~al.} (\bibinfo {collaboration} {ATLAS}),\ }\href {\doibase
  10.1007/JHEP11(2014)088} {\bibfield  {journal} {\bibinfo  {journal} {JHEP}\
  }\textbf {\bibinfo {volume} {11}},\ \bibinfo {pages} {088} (\bibinfo {year}
  {2014})},\ \Eprint {http://arxiv.org/abs/1409.0746} {arXiv:1409.0746
  [hep-ex]} \BibitemShut {NoStop}%
\bibitem [{\citenamefont {Aad}\ \emph {et~al.}(2016)\citenamefont {Aad} \emph
  {et~al.}}]{Aad:2015sms}%
  \BibitemOpen
  \bibfield  {author} {\bibinfo {author} {\bibfnamefont {G.}~\bibnamefont
  {Aad}} \emph {et~al.} (\bibinfo {collaboration} {ATLAS}),\ }\href {\doibase
  10.1007/JHEP02(2016)062} {\bibfield  {journal} {\bibinfo  {journal} {JHEP}\
  }\textbf {\bibinfo {volume} {02}},\ \bibinfo {pages} {062} (\bibinfo {year}
  {2016})},\ \Eprint {http://arxiv.org/abs/1511.05542} {arXiv:1511.05542
  [hep-ex]} \BibitemShut {NoStop}%
\bibitem [{\citenamefont {Khachatryan}\ \emph {et~al.}(2016)\citenamefont
  {Khachatryan} \emph {et~al.}}]{Khachatryan:2015wka}%
  \BibitemOpen
  \bibfield  {author} {\bibinfo {author} {\bibfnamefont {V.}~\bibnamefont
  {Khachatryan}} \emph {et~al.} (\bibinfo {collaboration} {CMS}),\ }\href
  {\doibase 10.1016/j.physletb.2015.10.067} {\bibfield  {journal} {\bibinfo
  {journal} {Phys. Lett.}\ }\textbf {\bibinfo {volume} {B752}},\ \bibinfo
  {pages} {146} (\bibinfo {year} {2016})},\ \Eprint
  {http://arxiv.org/abs/1506.00424} {arXiv:1506.00424 [hep-ex]} \BibitemShut
  {NoStop}%
\bibitem [{\citenamefont {collaboration}(2017)}]{ATLAS:2017dnw}%
  \BibitemOpen
  \bibfield  {author} {\bibinfo {author} {\bibfnamefont {T.~A.}\ \bibnamefont
  {collaboration}} (\bibinfo {collaboration} {ATLAS}),\ }\href@noop {} {\
  (\bibinfo {year} {2017})}\BibitemShut {NoStop}%
\bibitem [{CMS(2017)}]{CMS:2017odo}%
  \BibitemOpen
  \href {http://cds.cern.ch/record/2273394} {\emph {\bibinfo {title} {{Search
  for supersymmetry in events with at least one soft lepton, low jet
  multiplicity, and missing transverse momentum in proton-proton collisions at
  $\sqrt{s}=13~\mathrm{TeV}$}}}},\ \bibinfo {type} {Tech. Rep.}\ \bibinfo
  {number} {CMS-PAS-SUS-16-052}\ (\bibinfo  {institution} {CERN},\ \bibinfo
  {address} {Geneva},\ \bibinfo {year} {2017})\BibitemShut {NoStop}%
\bibitem [{\citenamefont {Aad}\ \emph {et~al.}(2015{\natexlab{a}})\citenamefont
  {Aad} \emph {et~al.}}]{Aad:2015asa}%
  \BibitemOpen
  \bibfield  {author} {\bibinfo {author} {\bibfnamefont {G.}~\bibnamefont
  {Aad}} \emph {et~al.} (\bibinfo {collaboration} {ATLAS}),\ }\href {\doibase
  10.1016/j.physletb.2015.02.015} {\bibfield  {journal} {\bibinfo  {journal}
  {Phys. Lett.}\ }\textbf {\bibinfo {volume} {B743}},\ \bibinfo {pages} {15}
  (\bibinfo {year} {2015}{\natexlab{a}})},\ \Eprint
  {http://arxiv.org/abs/1501.04020} {arXiv:1501.04020 [hep-ex]} \BibitemShut
  {NoStop}%
\bibitem [{\citenamefont {Aaij}\ \emph
  {et~al.}(2017{\natexlab{a}})\citenamefont {Aaij} \emph
  {et~al.}}]{Aaij:2016xmb}%
  \BibitemOpen
  \bibfield  {author} {\bibinfo {author} {\bibfnamefont {R.}~\bibnamefont
  {Aaij}} \emph {et~al.} (\bibinfo {collaboration} {LHCb}),\ }\href {\doibase
  10.1140/epjc/s10052-017-4744-6} {\bibfield  {journal} {\bibinfo  {journal}
  {Eur. Phys. J.}\ }\textbf {\bibinfo {volume} {C77}},\ \bibinfo {pages} {224}
  (\bibinfo {year} {2017}{\natexlab{a}})},\ \Eprint
  {http://arxiv.org/abs/1612.00945} {arXiv:1612.00945 [hep-ex]} \BibitemShut
  {NoStop}%
\bibitem [{\citenamefont {Holdom}(1986)}]{Holdom:1985ag}%
  \BibitemOpen
  \bibfield  {author} {\bibinfo {author} {\bibfnamefont {B.}~\bibnamefont
  {Holdom}},\ }\href {\doibase 10.1016/0370-2693(86)91377-8} {\bibfield
  {journal} {\bibinfo  {journal} {Phys. Lett.}\ }\textbf {\bibinfo {volume}
  {166B}},\ \bibinfo {pages} {196} (\bibinfo {year} {1986})}\BibitemShut
  {NoStop}%
\bibitem [{\citenamefont {Galison}\ and\ \citenamefont
  {Manohar}(1984)}]{Galison:1983pa}%
  \BibitemOpen
  \bibfield  {author} {\bibinfo {author} {\bibfnamefont {P.}~\bibnamefont
  {Galison}}\ and\ \bibinfo {author} {\bibfnamefont {A.}~\bibnamefont
  {Manohar}},\ }\href {\doibase 10.1016/0370-2693(84)91161-4} {\bibfield
  {journal} {\bibinfo  {journal} {Phys. Lett.}\ }\textbf {\bibinfo {volume}
  {136B}},\ \bibinfo {pages} {279} (\bibinfo {year} {1984})}\BibitemShut
  {NoStop}%
\bibitem [{\citenamefont {Hoenig}\ \emph {et~al.}(2014)\citenamefont {Hoenig},
  \citenamefont {Samach},\ and\ \citenamefont {Tucker-Smith}}]{Hoenig:2014dsa}%
  \BibitemOpen
  \bibfield  {author} {\bibinfo {author} {\bibfnamefont {I.}~\bibnamefont
  {Hoenig}}, \bibinfo {author} {\bibfnamefont {G.}~\bibnamefont {Samach}}, \
  and\ \bibinfo {author} {\bibfnamefont {D.}~\bibnamefont {Tucker-Smith}},\
  }\href {\doibase 10.1103/PhysRevD.90.075016} {\bibfield  {journal} {\bibinfo
  {journal} {Phys. Rev.}\ }\textbf {\bibinfo {volume} {D90}},\ \bibinfo {pages}
  {075016} (\bibinfo {year} {2014})},\ \Eprint {http://arxiv.org/abs/1408.1075}
  {arXiv:1408.1075 [hep-ph]} \BibitemShut {NoStop}%
\bibitem [{\citenamefont {Ilten}\ \emph {et~al.}(2016)\citenamefont {Ilten},
  \citenamefont {Soreq}, \citenamefont {Thaler}, \citenamefont {Williams},\
  and\ \citenamefont {Xue}}]{Ilten:2016tkc}%
  \BibitemOpen
  \bibfield  {author} {\bibinfo {author} {\bibfnamefont {P.}~\bibnamefont
  {Ilten}}, \bibinfo {author} {\bibfnamefont {Y.}~\bibnamefont {Soreq}},
  \bibinfo {author} {\bibfnamefont {J.}~\bibnamefont {Thaler}}, \bibinfo
  {author} {\bibfnamefont {M.}~\bibnamefont {Williams}}, \ and\ \bibinfo
  {author} {\bibfnamefont {W.}~\bibnamefont {Xue}},\ }\href {\doibase
  10.1103/PhysRevLett.116.251803} {\bibfield  {journal} {\bibinfo  {journal}
  {Phys. Rev. Lett.}\ }\textbf {\bibinfo {volume} {116}},\ \bibinfo {pages}
  {251803} (\bibinfo {year} {2016})},\ \Eprint
  {http://arxiv.org/abs/1603.08926} {arXiv:1603.08926 [hep-ph]} \BibitemShut
  {NoStop}%
\bibitem [{\citenamefont {Elahi}\ and\ \citenamefont
  {Martin}(2016)}]{Elahi:2015vzh}%
  \BibitemOpen
  \bibfield  {author} {\bibinfo {author} {\bibfnamefont {F.}~\bibnamefont
  {Elahi}}\ and\ \bibinfo {author} {\bibfnamefont {A.}~\bibnamefont {Martin}},\
  }\href {\doibase 10.1103/PhysRevD.93.015022} {\bibfield  {journal} {\bibinfo
  {journal} {Phys. Rev.}\ }\textbf {\bibinfo {volume} {D93}},\ \bibinfo {pages}
  {015022} (\bibinfo {year} {2016})},\ \Eprint
  {http://arxiv.org/abs/1511.04107} {arXiv:1511.04107 [hep-ph]} \BibitemShut
  {NoStop}%
\bibitem [{\citenamefont {Davoudiasl}\ \emph {et~al.}(2013)\citenamefont
  {Davoudiasl}, \citenamefont {Lee}, \citenamefont {Lewis},\ and\ \citenamefont
  {Marciano}}]{Davoudiasl:2013aya}%
  \BibitemOpen
  \bibfield  {author} {\bibinfo {author} {\bibfnamefont {H.}~\bibnamefont
  {Davoudiasl}}, \bibinfo {author} {\bibfnamefont {H.-S.}\ \bibnamefont {Lee}},
  \bibinfo {author} {\bibfnamefont {I.}~\bibnamefont {Lewis}}, \ and\ \bibinfo
  {author} {\bibfnamefont {W.~J.}\ \bibnamefont {Marciano}},\ }\href {\doibase
  10.1103/PhysRevD.88.015022} {\bibfield  {journal} {\bibinfo  {journal} {Phys.
  Rev.}\ }\textbf {\bibinfo {volume} {D88}},\ \bibinfo {pages} {015022}
  (\bibinfo {year} {2013})},\ \Eprint {http://arxiv.org/abs/1304.4935}
  {arXiv:1304.4935 [hep-ph]} \BibitemShut {NoStop}%
\bibitem [{\citenamefont {Curtin}\ \emph {et~al.}(2015)\citenamefont {Curtin},
  \citenamefont {Essig}, \citenamefont {Gori},\ and\ \citenamefont
  {Shelton}}]{Curtin:2014cca}%
  \BibitemOpen
  \bibfield  {author} {\bibinfo {author} {\bibfnamefont {D.}~\bibnamefont
  {Curtin}}, \bibinfo {author} {\bibfnamefont {R.}~\bibnamefont {Essig}},
  \bibinfo {author} {\bibfnamefont {S.}~\bibnamefont {Gori}}, \ and\ \bibinfo
  {author} {\bibfnamefont {J.}~\bibnamefont {Shelton}},\ }\href {\doibase
  10.1007/JHEP02(2015)157} {\bibfield  {journal} {\bibinfo  {journal} {JHEP}\
  }\textbf {\bibinfo {volume} {02}},\ \bibinfo {pages} {157} (\bibinfo {year}
  {2015})},\ \Eprint {http://arxiv.org/abs/1412.0018} {arXiv:1412.0018
  [hep-ph]} \BibitemShut {NoStop}%
\bibitem [{\citenamefont {Chou}\ \emph {et~al.}(2017)\citenamefont {Chou},
  \citenamefont {Curtin},\ and\ \citenamefont {Lubatti}}]{Chou:2016lxi}%
  \BibitemOpen
  \bibfield  {author} {\bibinfo {author} {\bibfnamefont {J.~P.}\ \bibnamefont
  {Chou}}, \bibinfo {author} {\bibfnamefont {D.}~\bibnamefont {Curtin}}, \ and\
  \bibinfo {author} {\bibfnamefont {H.~J.}\ \bibnamefont {Lubatti}},\ }\href
  {\doibase 10.1016/j.physletb.2017.01.043} {\bibfield  {journal} {\bibinfo
  {journal} {Phys. Lett.}\ }\textbf {\bibinfo {volume} {B767}},\ \bibinfo
  {pages} {29} (\bibinfo {year} {2017})},\ \Eprint
  {http://arxiv.org/abs/1606.06298} {arXiv:1606.06298 [hep-ph]} \BibitemShut
  {NoStop}%
\bibitem [{\citenamefont {Liu}\ \emph {et~al.}(2017)\citenamefont {Liu},
  \citenamefont {Wang},\ and\ \citenamefont {Yu}}]{Liu:2017lpo}%
  \BibitemOpen
  \bibfield  {author} {\bibinfo {author} {\bibfnamefont {J.}~\bibnamefont
  {Liu}}, \bibinfo {author} {\bibfnamefont {X.-P.}\ \bibnamefont {Wang}}, \
  and\ \bibinfo {author} {\bibfnamefont {F.}~\bibnamefont {Yu}},\ }\href
  {\doibase 10.1007/JHEP06(2017)077} {\bibfield  {journal} {\bibinfo  {journal}
  {JHEP}\ }\textbf {\bibinfo {volume} {06}},\ \bibinfo {pages} {077} (\bibinfo
  {year} {2017})},\ \Eprint {http://arxiv.org/abs/1704.00730} {arXiv:1704.00730
  [hep-ph]} \BibitemShut {NoStop}%
\bibitem [{\citenamefont {Lee}\ and\ \citenamefont
  {Weinberg}(1977)}]{Lee:1977ua}%
  \BibitemOpen
  \bibfield  {author} {\bibinfo {author} {\bibfnamefont {B.~W.}\ \bibnamefont
  {Lee}}\ and\ \bibinfo {author} {\bibfnamefont {S.}~\bibnamefont {Weinberg}},\
  }\href {\doibase 10.1103/PhysRevLett.39.165} {\bibfield  {journal} {\bibinfo
  {journal} {Phys. Rev. Lett.}\ }\textbf {\bibinfo {volume} {39}},\ \bibinfo
  {pages} {165} (\bibinfo {year} {1977})}\BibitemShut {NoStop}%
\bibitem [{\citenamefont {Pospelov}\ \emph {et~al.}(2008)\citenamefont
  {Pospelov}, \citenamefont {Ritz},\ and\ \citenamefont
  {Voloshin}}]{Pospelov:2007mp}%
  \BibitemOpen
  \bibfield  {author} {\bibinfo {author} {\bibfnamefont {M.}~\bibnamefont
  {Pospelov}}, \bibinfo {author} {\bibfnamefont {A.}~\bibnamefont {Ritz}}, \
  and\ \bibinfo {author} {\bibfnamefont {M.~B.}\ \bibnamefont {Voloshin}},\
  }\href {\doibase 10.1016/j.physletb.2008.02.052} {\bibfield  {journal}
  {\bibinfo  {journal} {Phys. Lett.}\ }\textbf {\bibinfo {volume} {B662}},\
  \bibinfo {pages} {53} (\bibinfo {year} {2008})},\ \Eprint
  {http://arxiv.org/abs/0711.4866} {arXiv:0711.4866 [hep-ph]} \BibitemShut
  {NoStop}%
\bibitem [{\citenamefont {Cvetic}\ and\ \citenamefont
  {Langacker}(1996)}]{Cvetic:1996mf}%
  \BibitemOpen
  \bibfield  {author} {\bibinfo {author} {\bibfnamefont {M.}~\bibnamefont
  {Cvetic}}\ and\ \bibinfo {author} {\bibfnamefont {P.}~\bibnamefont
  {Langacker}},\ }\href {\doibase 10.1142/S0217732396001260} {\bibfield
  {journal} {\bibinfo  {journal} {Mod. Phys. Lett.}\ }\textbf {\bibinfo
  {volume} {A11}},\ \bibinfo {pages} {1247} (\bibinfo {year} {1996})},\ \Eprint
  {http://arxiv.org/abs/hep-ph/9602424} {arXiv:hep-ph/9602424 [hep-ph]}
  \BibitemShut {NoStop}%
\bibitem [{\citenamefont {Abel}\ \emph {et~al.}(2008)\citenamefont {Abel},
  \citenamefont {Goodsell}, \citenamefont {Jaeckel}, \citenamefont {Khoze},\
  and\ \citenamefont {Ringwald}}]{Abel:2008ai}%
  \BibitemOpen
  \bibfield  {author} {\bibinfo {author} {\bibfnamefont {S.~A.}\ \bibnamefont
  {Abel}}, \bibinfo {author} {\bibfnamefont {M.~D.}\ \bibnamefont {Goodsell}},
  \bibinfo {author} {\bibfnamefont {J.}~\bibnamefont {Jaeckel}}, \bibinfo
  {author} {\bibfnamefont {V.~V.}\ \bibnamefont {Khoze}}, \ and\ \bibinfo
  {author} {\bibfnamefont {A.}~\bibnamefont {Ringwald}},\ }\href {\doibase
  10.1088/1126-6708/2008/07/124} {\bibfield  {journal} {\bibinfo  {journal}
  {JHEP}\ }\textbf {\bibinfo {volume} {07}},\ \bibinfo {pages} {124} (\bibinfo
  {year} {2008})},\ \Eprint {http://arxiv.org/abs/0803.1449} {arXiv:0803.1449
  [hep-ph]} \BibitemShut {NoStop}%
\bibitem [{\citenamefont {Goodsell}\ \emph {et~al.}(2009)\citenamefont
  {Goodsell}, \citenamefont {Jaeckel}, \citenamefont {Redondo},\ and\
  \citenamefont {Ringwald}}]{Goodsell:2009xc}%
  \BibitemOpen
  \bibfield  {author} {\bibinfo {author} {\bibfnamefont {M.}~\bibnamefont
  {Goodsell}}, \bibinfo {author} {\bibfnamefont {J.}~\bibnamefont {Jaeckel}},
  \bibinfo {author} {\bibfnamefont {J.}~\bibnamefont {Redondo}}, \ and\
  \bibinfo {author} {\bibfnamefont {A.}~\bibnamefont {Ringwald}},\ }\href
  {\doibase 10.1088/1126-6708/2009/11/027} {\bibfield  {journal} {\bibinfo
  {journal} {JHEP}\ }\textbf {\bibinfo {volume} {11}},\ \bibinfo {pages} {027}
  (\bibinfo {year} {2009})},\ \Eprint {http://arxiv.org/abs/0909.0515}
  {arXiv:0909.0515 [hep-ph]} \BibitemShut {NoStop}%
\bibitem [{\citenamefont {Lees}\ \emph {et~al.}(2014)\citenamefont {Lees} \emph
  {et~al.}}]{Lees:2014xha}%
  \BibitemOpen
  \bibfield  {author} {\bibinfo {author} {\bibfnamefont {J.~P.}\ \bibnamefont
  {Lees}} \emph {et~al.} (\bibinfo {collaboration} {BaBar}),\ }\href {\doibase
  10.1103/PhysRevLett.113.201801} {\bibfield  {journal} {\bibinfo  {journal}
  {Phys. Rev. Lett.}\ }\textbf {\bibinfo {volume} {113}},\ \bibinfo {pages}
  {201801} (\bibinfo {year} {2014})},\ \Eprint {http://arxiv.org/abs/1406.2980}
  {arXiv:1406.2980 [hep-ex]} \BibitemShut {NoStop}%
\bibitem [{\citenamefont {Lees}\ \emph {et~al.}(2012)\citenamefont {Lees} \emph
  {et~al.}}]{Lees:2012ra}%
  \BibitemOpen
  \bibfield  {author} {\bibinfo {author} {\bibfnamefont {J.~P.}\ \bibnamefont
  {Lees}} \emph {et~al.} (\bibinfo {collaboration} {BaBar}),\ }\href {\doibase
  10.1103/PhysRevLett.108.211801} {\bibfield  {journal} {\bibinfo  {journal}
  {Phys. Rev. Lett.}\ }\textbf {\bibinfo {volume} {108}},\ \bibinfo {pages}
  {211801} (\bibinfo {year} {2012})},\ \Eprint {http://arxiv.org/abs/1202.1313}
  {arXiv:1202.1313 [hep-ex]} \BibitemShut {NoStop}%
\bibitem [{\citenamefont {Jaegle}(2015)}]{TheBelle:2015mwa}%
  \BibitemOpen
  \bibfield  {author} {\bibinfo {author} {\bibfnamefont {I.}~\bibnamefont
  {Jaegle}} (\bibinfo {collaboration} {Belle}),\ }\href {\doibase
  10.1103/PhysRevLett.114.211801} {\bibfield  {journal} {\bibinfo  {journal}
  {Phys. Rev. Lett.}\ }\textbf {\bibinfo {volume} {114}},\ \bibinfo {pages}
  {211801} (\bibinfo {year} {2015})},\ \Eprint
  {http://arxiv.org/abs/1502.00084} {arXiv:1502.00084 [hep-ex]} \BibitemShut
  {NoStop}%
\bibitem [{\citenamefont {Hook}\ \emph {et~al.}(2011)\citenamefont {Hook},
  \citenamefont {Izaguirre},\ and\ \citenamefont {Wacker}}]{Hook:2010tw}%
  \BibitemOpen
  \bibfield  {author} {\bibinfo {author} {\bibfnamefont {A.}~\bibnamefont
  {Hook}}, \bibinfo {author} {\bibfnamefont {E.}~\bibnamefont {Izaguirre}}, \
  and\ \bibinfo {author} {\bibfnamefont {J.~G.}\ \bibnamefont {Wacker}},\
  }\href {\doibase 10.1155/2011/859762} {\bibfield  {journal} {\bibinfo
  {journal} {Adv. High Energy Phys.}\ }\textbf {\bibinfo {volume} {2011}},\
  \bibinfo {pages} {859762} (\bibinfo {year} {2011})},\ \Eprint
  {http://arxiv.org/abs/1006.0973} {arXiv:1006.0973 [hep-ph]} \BibitemShut
  {NoStop}%
\bibitem [{\citenamefont {Chatrchyan}\ \emph
  {et~al.}(2013{\natexlab{a}})\citenamefont {Chatrchyan} \emph
  {et~al.}}]{Chatrchyan:2013tia}%
  \BibitemOpen
  \bibfield  {author} {\bibinfo {author} {\bibfnamefont {S.}~\bibnamefont
  {Chatrchyan}} \emph {et~al.} (\bibinfo {collaboration} {CMS}),\ }\href
  {\doibase 10.1007/JHEP12(2013)030} {\bibfield  {journal} {\bibinfo  {journal}
  {JHEP}\ }\textbf {\bibinfo {volume} {12}},\ \bibinfo {pages} {030} (\bibinfo
  {year} {2013}{\natexlab{a}})},\ \Eprint {http://arxiv.org/abs/1310.7291}
  {arXiv:1310.7291 [hep-ex]} \BibitemShut {NoStop}%
\bibitem [{\citenamefont {Aaij}\ \emph
  {et~al.}(2017{\natexlab{b}})\citenamefont {Aaij} \emph
  {et~al.}}]{Aaij:2017rft}%
  \BibitemOpen
  \bibfield  {author} {\bibinfo {author} {\bibfnamefont {R.}~\bibnamefont
  {Aaij}} \emph {et~al.} (\bibinfo {collaboration} {LHCb}),\ }\href@noop {} {\
  (\bibinfo {year} {2017}{\natexlab{b}})},\ \Eprint
  {http://arxiv.org/abs/1710.02867} {arXiv:1710.02867 [hep-ex]} \BibitemShut
  {NoStop}%
\bibitem [{\citenamefont {Aad}\ \emph {et~al.}(2015{\natexlab{b}})\citenamefont
  {Aad} \emph {et~al.}}]{Aad:2015sva}%
  \BibitemOpen
  \bibfield  {author} {\bibinfo {author} {\bibfnamefont {G.}~\bibnamefont
  {Aad}} \emph {et~al.} (\bibinfo {collaboration} {ATLAS}),\ }\href {\doibase
  10.1103/PhysRevD.92.092001} {\bibfield  {journal} {\bibinfo  {journal} {Phys.
  Rev.}\ }\textbf {\bibinfo {volume} {D92}},\ \bibinfo {pages} {092001}
  (\bibinfo {year} {2015}{\natexlab{b}})},\ \Eprint
  {http://arxiv.org/abs/1505.07645} {arXiv:1505.07645 [hep-ex]} \BibitemShut
  {NoStop}%
\bibitem [{ATL(2017)}]{ATL-DAQ-PUB-2017-001}%
  \BibitemOpen
  \href {https://cds.cern.ch/record/2242069} {\emph {\bibinfo {title} {{Trigger
  Menu in 2016}}}},\ \bibinfo {type} {Tech. Rep.}\ \bibinfo {number}
  {ATL-DAQ-PUB-2017-001}\ (\bibinfo  {institution} {CERN},\ \bibinfo {address}
  {Geneva},\ \bibinfo {year} {2017})\BibitemShut {NoStop}%
\bibitem [{\citenamefont {Khachatryan}\ \emph
  {et~al.}(2015{\natexlab{a}})\citenamefont {Khachatryan} \emph
  {et~al.}}]{CMS:2014xja}%
  \BibitemOpen
  \bibfield  {author} {\bibinfo {author} {\bibfnamefont {V.}~\bibnamefont
  {Khachatryan}} \emph {et~al.} (\bibinfo {collaboration} {CMS}),\ }\href
  {\doibase 10.1016/j.physletb.2016.04.010, 10.1016/j.physletb.2014.11.059}
  {\bibfield  {journal} {\bibinfo  {journal} {Phys. Lett.}\ }\textbf {\bibinfo
  {volume} {B740}},\ \bibinfo {pages} {250} (\bibinfo {year}
  {2015}{\natexlab{a}})},\ \bibinfo {note} {[erratum: Phys.
  Lett.B757,569(2016)]},\ \Eprint {http://arxiv.org/abs/1406.0113}
  {arXiv:1406.0113 [hep-ex]} \BibitemShut {NoStop}%
\bibitem [{\citenamefont {Khachatryan}\ \emph {et~al.}(2014)\citenamefont
  {Khachatryan} \emph {et~al.}}]{Khachatryan:2014qwa}%
  \BibitemOpen
  \bibfield  {author} {\bibinfo {author} {\bibfnamefont {V.}~\bibnamefont
  {Khachatryan}} \emph {et~al.} (\bibinfo {collaboration} {CMS}),\ }\href
  {\doibase 10.1140/epjc/s10052-014-3036-7} {\bibfield  {journal} {\bibinfo
  {journal} {Eur. Phys. J.}\ }\textbf {\bibinfo {volume} {C74}},\ \bibinfo
  {pages} {3036} (\bibinfo {year} {2014})},\ \Eprint
  {http://arxiv.org/abs/1405.7570} {arXiv:1405.7570 [hep-ex]} \BibitemShut
  {NoStop}%
\bibitem [{\citenamefont {Sirunyan}\ \emph {et~al.}(2017)\citenamefont
  {Sirunyan} \emph {et~al.}}]{Sirunyan:2017lae}%
  \BibitemOpen
  \bibfield  {author} {\bibinfo {author} {\bibfnamefont {A.~M.}\ \bibnamefont
  {Sirunyan}} \emph {et~al.} (\bibinfo {collaboration} {CMS}),\ }\href@noop {}
  {\  (\bibinfo {year} {2017})},\ \Eprint {http://arxiv.org/abs/1709.05406}
  {arXiv:1709.05406 [hep-ex]} \BibitemShut {NoStop}%
\bibitem [{\citenamefont {Alwall}\ \emph {et~al.}(2014)\citenamefont {Alwall},
  \citenamefont {Frederix}, \citenamefont {Frixione}, \citenamefont {Hirschi},
  \citenamefont {Maltoni}, \citenamefont {Mattelaer}, \citenamefont {Shao},
  \citenamefont {Stelzer}, \citenamefont {Torrielli},\ and\ \citenamefont
  {Zaro}}]{Alwall:2014hca}%
  \BibitemOpen
  \bibfield  {author} {\bibinfo {author} {\bibfnamefont {J.}~\bibnamefont
  {Alwall}}, \bibinfo {author} {\bibfnamefont {R.}~\bibnamefont {Frederix}},
  \bibinfo {author} {\bibfnamefont {S.}~\bibnamefont {Frixione}}, \bibinfo
  {author} {\bibfnamefont {V.}~\bibnamefont {Hirschi}}, \bibinfo {author}
  {\bibfnamefont {F.}~\bibnamefont {Maltoni}}, \bibinfo {author} {\bibfnamefont
  {O.}~\bibnamefont {Mattelaer}}, \bibinfo {author} {\bibfnamefont {H.~S.}\
  \bibnamefont {Shao}}, \bibinfo {author} {\bibfnamefont {T.}~\bibnamefont
  {Stelzer}}, \bibinfo {author} {\bibfnamefont {P.}~\bibnamefont {Torrielli}},
  \ and\ \bibinfo {author} {\bibfnamefont {M.}~\bibnamefont {Zaro}},\ }\href
  {\doibase 10.1007/JHEP07(2014)079} {\bibfield  {journal} {\bibinfo  {journal}
  {JHEP}\ }\textbf {\bibinfo {volume} {07}},\ \bibinfo {pages} {079} (\bibinfo
  {year} {2014})},\ \Eprint {http://arxiv.org/abs/1405.0301} {arXiv:1405.0301
  [hep-ph]} \BibitemShut {NoStop}%
\bibitem [{\citenamefont {Alwall}\ \emph
  {et~al.}(2009{\natexlab{b}})\citenamefont {Alwall}, \citenamefont
  {de~Visscher},\ and\ \citenamefont {Maltoni}}]{Alwall:2008qv}%
  \BibitemOpen
  \bibfield  {author} {\bibinfo {author} {\bibfnamefont {J.}~\bibnamefont
  {Alwall}}, \bibinfo {author} {\bibfnamefont {S.}~\bibnamefont {de~Visscher}},
  \ and\ \bibinfo {author} {\bibfnamefont {F.}~\bibnamefont {Maltoni}},\ }\href
  {\doibase 10.1088/1126-6708/2009/02/017} {\bibfield  {journal} {\bibinfo
  {journal} {JHEP}\ }\textbf {\bibinfo {volume} {02}},\ \bibinfo {pages} {017}
  (\bibinfo {year} {2009}{\natexlab{b}})},\ \Eprint
  {http://arxiv.org/abs/0810.5350} {arXiv:0810.5350 [hep-ph]} \BibitemShut
  {NoStop}%
\bibitem [{\citenamefont {Sjostrand}\ \emph {et~al.}(2006)\citenamefont
  {Sjostrand}, \citenamefont {Mrenna},\ and\ \citenamefont
  {Skands}}]{Sjostrand:2006za}%
  \BibitemOpen
  \bibfield  {author} {\bibinfo {author} {\bibfnamefont {T.}~\bibnamefont
  {Sjostrand}}, \bibinfo {author} {\bibfnamefont {S.}~\bibnamefont {Mrenna}}, \
  and\ \bibinfo {author} {\bibfnamefont {P.~Z.}\ \bibnamefont {Skands}},\
  }\href {\doibase 10.1088/1126-6708/2006/05/026} {\bibfield  {journal}
  {\bibinfo  {journal} {JHEP}\ }\textbf {\bibinfo {volume} {05}},\ \bibinfo
  {pages} {026} (\bibinfo {year} {2006})},\ \Eprint
  {http://arxiv.org/abs/hep-ph/0603175} {arXiv:hep-ph/0603175 [hep-ph]}
  \BibitemShut {NoStop}%
\bibitem [{\citenamefont {Sjostrand}\ \emph {et~al.}(2008)\citenamefont
  {Sjostrand}, \citenamefont {Mrenna},\ and\ \citenamefont
  {Skands}}]{Sjostrand:2007gs}%
  \BibitemOpen
  \bibfield  {author} {\bibinfo {author} {\bibfnamefont {T.}~\bibnamefont
  {Sjostrand}}, \bibinfo {author} {\bibfnamefont {S.}~\bibnamefont {Mrenna}}, \
  and\ \bibinfo {author} {\bibfnamefont {P.~Z.}\ \bibnamefont {Skands}},\
  }\href {\doibase 10.1016/j.cpc.2008.01.036} {\bibfield  {journal} {\bibinfo
  {journal} {Comput. Phys. Commun.}\ }\textbf {\bibinfo {volume} {178}},\
  \bibinfo {pages} {852} (\bibinfo {year} {2008})},\ \Eprint
  {http://arxiv.org/abs/0710.3820} {arXiv:0710.3820 [hep-ph]} \BibitemShut
  {NoStop}%
\bibitem [{\citenamefont {Sjostrand}\ \emph {et~al.}(2015)\citenamefont
  {Sjostrand}, \citenamefont {Ask}, \citenamefont {Christiansen}, \citenamefont
  {Corke}, \citenamefont {Desai}, \citenamefont {Ilten}, \citenamefont
  {Mrenna}, \citenamefont {Prestel}, \citenamefont {Rasmussen},\ and\
  \citenamefont {Skands}}]{Sjostrand:2014zea}%
  \BibitemOpen
  \bibfield  {author} {\bibinfo {author} {\bibfnamefont {T.}~\bibnamefont
  {Sjostrand}}, \bibinfo {author} {\bibfnamefont {S.}~\bibnamefont {Ask}},
  \bibinfo {author} {\bibfnamefont {J.~R.}\ \bibnamefont {Christiansen}},
  \bibinfo {author} {\bibfnamefont {R.}~\bibnamefont {Corke}}, \bibinfo
  {author} {\bibfnamefont {N.}~\bibnamefont {Desai}}, \bibinfo {author}
  {\bibfnamefont {P.}~\bibnamefont {Ilten}}, \bibinfo {author} {\bibfnamefont
  {S.}~\bibnamefont {Mrenna}}, \bibinfo {author} {\bibfnamefont
  {S.}~\bibnamefont {Prestel}}, \bibinfo {author} {\bibfnamefont {C.~O.}\
  \bibnamefont {Rasmussen}}, \ and\ \bibinfo {author} {\bibfnamefont {P.~Z.}\
  \bibnamefont {Skands}},\ }\href {\doibase 10.1016/j.cpc.2015.01.024}
  {\bibfield  {journal} {\bibinfo  {journal} {Comput. Phys. Commun.}\ }\textbf
  {\bibinfo {volume} {191}},\ \bibinfo {pages} {159} (\bibinfo {year}
  {2015})},\ \Eprint {http://arxiv.org/abs/1410.3012} {arXiv:1410.3012
  [hep-ph]} \BibitemShut {NoStop}%
\bibitem [{\citenamefont {Read}(2002)}]{Read:2002hq}%
  \BibitemOpen
  \bibfield  {author} {\bibinfo {author} {\bibfnamefont {A.~L.}\ \bibnamefont
  {Read}},\ }\bibfield  {booktitle} {\emph {\bibinfo {booktitle} {{Advanced
  Statistical Techniques in Particle Physics. Proceedings, Conference, Durham,
  UK, March 18-22, 2002}}},\ }\href {\doibase 10.1088/0954-3899/28/10/313}
  {\bibfield  {journal} {\bibinfo  {journal} {J. Phys.}\ }\textbf {\bibinfo
  {volume} {G28}},\ \bibinfo {pages} {2693} (\bibinfo {year} {2002})},\
  \bibinfo {note} {[,11(2002)]}\BibitemShut {NoStop}%
\bibitem [{\citenamefont {Chatrchyan}\ \emph
  {et~al.}(2013{\natexlab{b}})\citenamefont {Chatrchyan} \emph
  {et~al.}}]{Chatrchyan:2012cg}%
  \BibitemOpen
  \bibfield  {author} {\bibinfo {author} {\bibfnamefont {S.}~\bibnamefont
  {Chatrchyan}} \emph {et~al.} (\bibinfo {collaboration} {CMS}),\ }\href
  {\doibase 10.1016/j.physletb.2013.09.009} {\bibfield  {journal} {\bibinfo
  {journal} {Phys. Lett.}\ }\textbf {\bibinfo {volume} {B726}},\ \bibinfo
  {pages} {564} (\bibinfo {year} {2013}{\natexlab{b}})},\ \Eprint
  {http://arxiv.org/abs/1210.7619} {arXiv:1210.7619 [hep-ex]} \BibitemShut
  {NoStop}%
\bibitem [{\citenamefont {Aad}\ \emph {et~al.}(2015{\natexlab{c}})\citenamefont
  {Aad} \emph {et~al.}}]{Aad:2015rba}%
  \BibitemOpen
  \bibfield  {author} {\bibinfo {author} {\bibfnamefont {G.}~\bibnamefont
  {Aad}} \emph {et~al.} (\bibinfo {collaboration} {ATLAS}),\ }\href {\doibase
  10.1103/PhysRevD.92.072004} {\bibfield  {journal} {\bibinfo  {journal} {Phys.
  Rev.}\ }\textbf {\bibinfo {volume} {D92}},\ \bibinfo {pages} {072004}
  (\bibinfo {year} {2015}{\natexlab{c}})},\ \Eprint
  {http://arxiv.org/abs/1504.05162} {arXiv:1504.05162 [hep-ex]} \BibitemShut
  {NoStop}%
\bibitem [{\citenamefont {Khachatryan}\ \emph
  {et~al.}(2015{\natexlab{b}})\citenamefont {Khachatryan} \emph
  {et~al.}}]{CMS:2014hka}%
  \BibitemOpen
  \bibfield  {author} {\bibinfo {author} {\bibfnamefont {V.}~\bibnamefont
  {Khachatryan}} \emph {et~al.} (\bibinfo {collaboration} {CMS}),\ }\href
  {\doibase 10.1103/PhysRevD.91.052012} {\bibfield  {journal} {\bibinfo
  {journal} {Phys. Rev.}\ }\textbf {\bibinfo {volume} {D91}},\ \bibinfo {pages}
  {052012} (\bibinfo {year} {2015}{\natexlab{b}})},\ \Eprint
  {http://arxiv.org/abs/1411.6977} {arXiv:1411.6977 [hep-ex]} \BibitemShut
  {NoStop}%
\end{thebibliography}%

\end{document}